\documentstyle[preprint,aps]{revtex}

\input epsf
\newcommand{\bobfig}[1]{
\makebox{\epsfysize=4.5mm \epsfbox{#1}}}
\newcommand{\bookfig}[5]{
\begin{figure}\centering\fbox{\epsfysize=#5cm \epsfbox{#1}}
\caption[#2]{#4}\label{#3}\end{figure}}

\tightenlines  

\begin{document}
\draft

\narrowtext
\title{Using the Hopf algebra structure of QFT in calculations}
\author{D.~Kreimer\cite{Author1}}
\address{Department of Physics, Mainz University, D-55099 Mainz, Germany}
\author{R.~Delbourgo\cite{Author2}}
\address{University of Tasmania, GPO Box 252-21, Hobart,
 Tasmania 7001, Australia}
\date{\today,hep-th/9903249}
\maketitle

\begin{abstract}
We employ the recently discovered Hopf algebra structure underlying perturbative
Quantum Field Theory to derive iterated integral representations
for Feynman diagrams. We give two applications: to massless Yukawa theory and
quantum electrodynamics in four dimensions.
\end{abstract}

\pacs{11.10.Gh, 11.15.Bt}  

\section{Introduction}
Quantum field theories (QFT) of interest to particle physicists
are plagued by UV divergences. Typically, we are confronted with
theories which favour local interactions, which means that we can
remove these divergences order by order in the loop expansion.
This iterative removal of divergences, all the while maintaining
the locality of the theory, has to fulfill certain combinatorial
properties, succinctly summarized by Zimmermann's forest formula.
It was a great achievement when the self-consistency of the whole
renormalization procedure was proven \cite{BPHZ}. While there is
little doubt that renormalization theory allows one to obtain
sensible answers from an a priori ill-defined theory, only
recently has it become clear that its underlying mathematical
structure is by no means accidental, for it relates QFT to basic
mathematical notions, well-known in low dimensional topology,
number theory \cite{hopf,CK,overl,DiDa,dkchen} and even in numerical
analysis \cite{RK}.

In this paper, we want to show to what extent the renormalization
of Feynman graphs can be incorporated in the language of
generalized iterated integrals, and how the intricacies of
full-fledged QFT, including spin and Lorentz structure, still can
be described in terms of such generalized iterated integrals in
the spirit of \cite{dkchen}.

We will begin by describing how the Hopf algebra structure,
associated with renormalization, can be related to rooted trees,
through the nature of the Feynman graphs. In that opening section
we also show how two kinds of renormalization schemes can be
connected by a convolution product; this convolution is the group
product naturally related to the Hopf algebra, and dutifully
delivers the renormalization group. The next section explains how
rooted trees acquire decorations, according to the topology of the
Feynman diagram, and how their evaluation is completely determined
by vertex weights and generalized tree factorials. The following
section concerns the representation of Feynman diagrams as
iterated one-dimensional integrals and how this can be tied to the
renormalization group, through appropriate choices of end-points.
Finally we treat the case of massless Yukawa theory and quantum
electrodynamics in four dimensions; the former sets the scene for
tackling scalar theories, while the latter contains the necessary
matrix generalization to spin, where the occurrence of
form-factors can complicate the argument. The paper ends with a
brief section embodying our conclusions and highlighting topics
for further development.

\section{Formulary for Renormalization}
In this section we will briefly describe the Hopf algebra structure of
renormalization. We start with a basic formulation directly on graphs
and then introduce the correspondence to rooted trees.

\subsection{Basics}
Let us consider a Feynman graph $\Gamma$ as a set, consisting
of vertices and edges of several sorts.
The edges correspond to various types of propagators,
representing inverse differential operators of free relativistic wave
equations with boundary conditions in accord with causality.
The vertices correspond to local interactions associated with a
Lagrangian which is a polynomial in fields and derivatives. In this
way, at the same time $\Gamma$ denotes a graph which we can draw on
paper as well as a unique analytical expression.

Regarding $\Gamma$ as a set of edges and vertices, we can formally consider
the power set of this set, and distinguish those elements of this power
set which themselves correspond to superficially divergent Feynman graphs.

Such a Feynman graph, together with all its superficially
divergent subgraphs, constitutes a Hopf algebra structure which is
isomorphic to a Hopf algebra of decorated rooted trees. This
isomorphism is described in Fig. \ref{b1} and  is detailed in
\cite{CK,overl}. (For the reader's convenience, we have added
an Appendix which summarizes the crucial points.)
By this isomorphism, the standard Hopf algebra operations on
rooted trees correspond to established notions of physicists.

Admissible cuts on the sum of rooted trees representing $\Gamma$
are in one-to-one correspondence with divergent subgraphs.  Those
are determined by  power-counting, thereby allowing detection of all
sectors in the analytic expression $\Gamma$ which correspond to
divergent subintegrations in their own right, while the forests of
renormalization theory then correspond to arbitrary cuts
\cite{hopf,CK,overl}.

This viewpoint leads to a Hopf algebra structure
\cite{hopf,CK,overl,DiDa,dkchen}, which we can succinctly
formulate on overall divergent Feynman graphs $\Gamma$ as
\begin{eqnarray}
\Delta[1] & = & 1\otimes 1\\
\Delta[\Gamma] & = & \sum_{\gamma{\subseteq\atop X}\Gamma}\gamma
\otimes\Gamma/\gamma\\
 S[1] & = & 1\\
S[\Gamma] & = & -\Gamma
-\sum_{\gamma{\subset\atop X}\Gamma}\gamma\otimes\Gamma/\gamma\\
 & = &
-\Gamma-\sum_{F\in {\cal F}}(-1)^{n_F}F\otimes\Gamma/F\\
\bar{e}[1] & = & 1\\
\bar{e}[\Gamma] & = & 0.
\end{eqnarray}
In this notation, the sum $\sum_{\gamma{\subseteq\atop X}\Gamma}$
is a sum over all superficially divergent subgraphs, including the
case $\gamma=\Gamma$ and $\gamma=\emptyset$. $\Gamma/\gamma$ is
obtained by  the contraction of $\gamma\subset\Gamma$ to a point
in $\Gamma$. The algebra element corresponding to the empty set
is, as usual, the unit $1$ of the algebra. Hence, the coproduct
determines all sectors in the graph which are ill-defined and
require renormalization. The sum $\sum_{\gamma{\subset\atop
X}\Gamma}$ is a similar sum, with the two cases $\gamma=\Gamma$
and $\gamma=\emptyset$ excluded.

${\cal F}$ is the set of all forests of a Feynman graph, and this
set is in one-to-one correspondence with all cuts at the rooted
tree $t_\Gamma$ representing the graph  $\Gamma$, while divergent
subgraphs are in one-to-one correspondence with all admissible
cuts, as already mentioned. $n_F$ is the number of boxes of the
forest $F$, or the number of elementary cuts under the above
correspondence \cite{CK}. Since formulae (1)-(8) are rather
abstract for the uninitiated reader, we give an illuminating
example in Fig. \ref{b1}; a proper derivation of these
correspondences can be found in \cite{hopf,CK,overl} while the
Appendix summarizes the crucial notions.


Hopf algebras, as embodied in (1)-(8) are Hopf algebras of rooted
trees for an appropriately chosen set of decorations \cite{overl}.
With this Hopf algebra structure, the naive renormalized  function
${\bf \Gamma}[\Gamma]$ associated with $\Gamma$ vanishes:
\begin{equation}
{\bf \Gamma}[\Gamma]:=m[(S\otimes id)\Delta[\Gamma]]=\bar{e}[\Gamma]=0.
\end{equation}
Spelling the formula out, one indeed finds that each term appears
twice with opposite signs. This must be so as the above formula
employs the graph itself as the counterterm of a divergent graph,
and hence eliminates all divergent graphs. Equation (8) represents a
well-defined renormalization scheme, but is not of much use in
practice. It has a few merits in the large $N_C$ expansion however
\cite{tHooft}.

Usually we want to subtract in accordance with certain renormalization
conditions, and hence our counterterms should surely remove divergences
of integrals but may not completely nullify all superficially divergent
bare Green functions. We have to choose non-trivial renormalization
schemes to obtain non-vanishing but finite results. The essential modification
is to allow for a map $R$ which modifies the bare unrenormalized Feynman
graph without changing its divergence. In \cite{dkchen} it was
shown that all renormalization schemes can be described on the
same footing, allowing for tree-dependent variations of
dimensionful parameters in bare Green functions.

\subsection{Schemes and scale transformations}
We can use the results of \cite{dkchen} to recast the behaviour
under a change of parameters and the dependence on renormalization
schemes in a nice manner. We first modify the antipode via a
renormalization map $R$, and in consequence get a finite
renormalized Green function ${\bf \Gamma}_R[\Gamma]$
\cite{hopf,CK,overl,DiDa,dkchen}. The $R$ dependent antipode, the
$Z$-factor $S_R[\Gamma]$ of a bare expression $\Gamma$ and the
renormalized Green function ${\bf \Gamma}_R[\Gamma]$ explicitly
read
\begin{eqnarray}
S_R[\Gamma] & = & R\left[-\Gamma -\sum_{\gamma{\subset\atop
X}\Gamma}S_R[\gamma]\Gamma/\gamma \right]\label{antip}\\
 & = &
R\left[ \sum_{F\in{\cal
F}}(-1)^{n_F+1}F_R(\gamma_F)\Gamma/\gamma_F\right]\\ {\bf
\Gamma}_R[\Gamma] & = & m[(S_R\otimes
id)\Delta[\Gamma]]\label{ren}\\
 & = & [id-R]\left(\sum_{F\in{\cal F}}(-1)^{n_F}F_R(\gamma_F)\Gamma/\gamma_F\right).
\end{eqnarray}
Here, $\gamma_F$ is the subgraph corresponding to the forest $F$,
and $F_R$ is the evaluation of this subgraph using $R$ to evaluate
boxes of forests corresponding to cuts, cf.~Fig. \ref{b2}.

Introducing the convolution product
$(\phi\star\psi)(\Gamma)=m[(\phi\otimes\psi)\Delta(\Gamma)]$ we
find
\begin{equation}
{\bf\Gamma}_R[\Gamma]=(S_R\star id)[\Gamma],
\end{equation}
which makes the renormalized Green function transparent as a ratio
of a bare Green function $\Gamma\equiv id(\Gamma)$
to a counterterm $S_R(\Gamma)$. Hence our
notation emphasizes the fact that the renormalized Green function
is a function of the bare Green function as well as the chosen
renormalization scheme. A detailed discussion of all its
properties can be found in \cite{dkchen}. Here we only remark that
the group structure embodied in the  renormalization group is a
direct consequence of the structure of a renormalized Green
function as a convolution product \cite{DiDa,dkchen}.

Of course, bare functions $\Gamma$ are parametrized by masses,
external momenta and regularization (cut-off, dimensional-regularization
scale $\mu$, etc.). We can vary such parameters. Essentially, the choice
of a renormalization scheme amounts to a choice of conditions for
external parameters. Hence, a renormalized Feynman graph is
determined by the choice of two sets of such parameters, one for
the bare diagrams, one for the counterterm diagrams. If we want to
highlight the dependence of the renormalized Green function on such
parameter sets $i,k$ say, we explicitly write ${\bf \Gamma}_{R_i,k}$;
this assigns to a Feynman graph a renormalized
expression with a specified set of dimensionful parameters $k$
used in the bare Green functions and a set of parameters $i$ used
for counterterms. For example, an on-shell renormalized off-shell
Green function will be calculated by using bare Feynman diagrams
with off-shell momenta and imposing on-shell conditions to fix them
in the counterterms.

Now, from \cite{dkchen}, we know that the most general renormalization
scheme, including for example minimal subtraction (MS), can be captured
by suitable modifications of these scales. There one discovers that
the change of schemes and scales can be succinctly summarized by
the following formula
\begin{equation}
{\bf \Gamma}_{R_i,k}(\Gamma)=[{\bf \Gamma}_{R_i,j}\star{\bf
\Gamma}_{R_j,k}](\Gamma).\label{chenij}
\end{equation}
In ${\bf \Gamma}_{R_i,j}$, the first subscript at the renormalization
map $R$ indicates the choice of scales for the counterterms, while
the second one indicates the set of external scales
used in the bare diagrams. Equation (\ref{chenij}) shows how an
intermediate set of scales $j$ allows for group-like
transformation laws. From this, one obtains renormalization group
equations, operator product expansions and cohomological
properties of renormalization \cite{dkchen} with ease. Depending
on the choice of parameters to be varied, one arrives at Wilson's
viewpoint by varying  a cut-off \cite{WI} or at the Callan-Symanzik
equation from a variation of the renormalization point. Details
will be given in future work.

\subsection{Example}
The following example, of a three-loop fermion self-energy
contribution in QED in four dimensions, will make the above ideas
more transparent and their implementation more concrete. We first
calculate the coproduct for the diagram:
\begin{eqnarray*}
\Delta\left[\bobfig{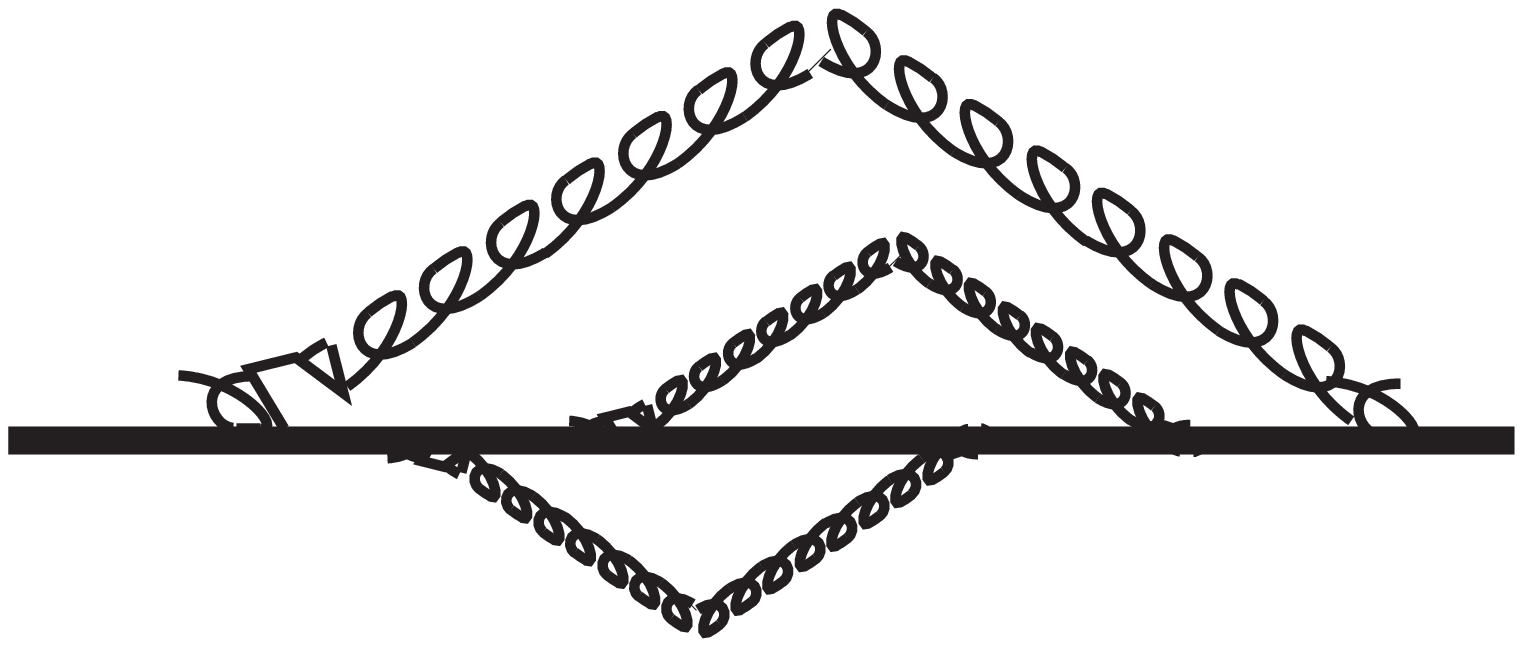}\right] & = &
\mbox{\bobfig{bob1.eps}}\otimes 1+1\otimes \mbox{\bobfig{bob1.eps}}
 +\mbox{\bobfig{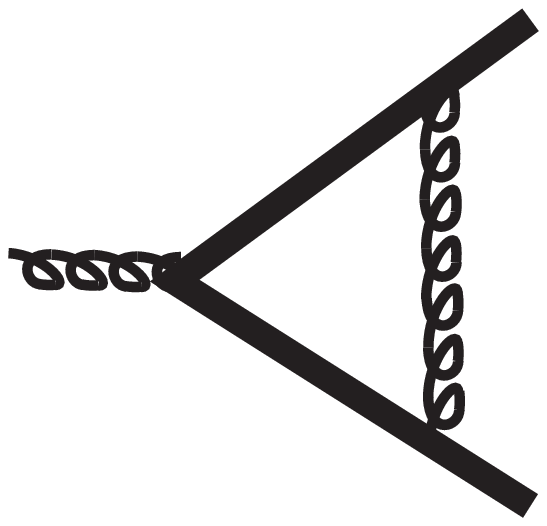}}\otimes\mbox{\bobfig{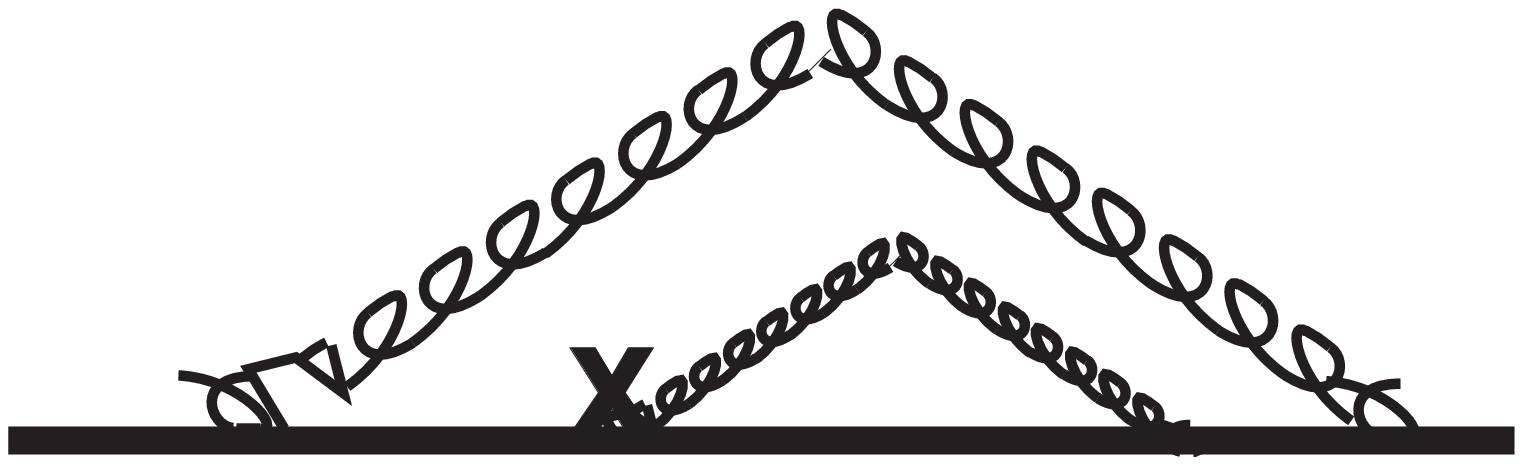}}\\
 & & +\mbox{\bobfig{bob4.eps}}\otimes\mbox{\bobfig{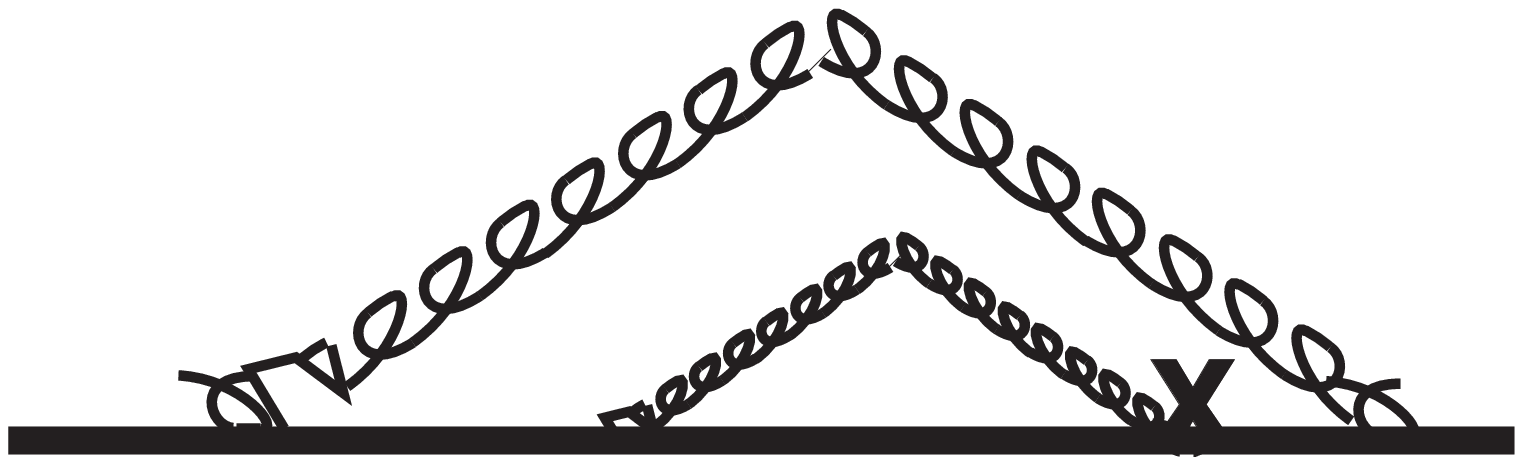}}
  +\mbox{\bobfig{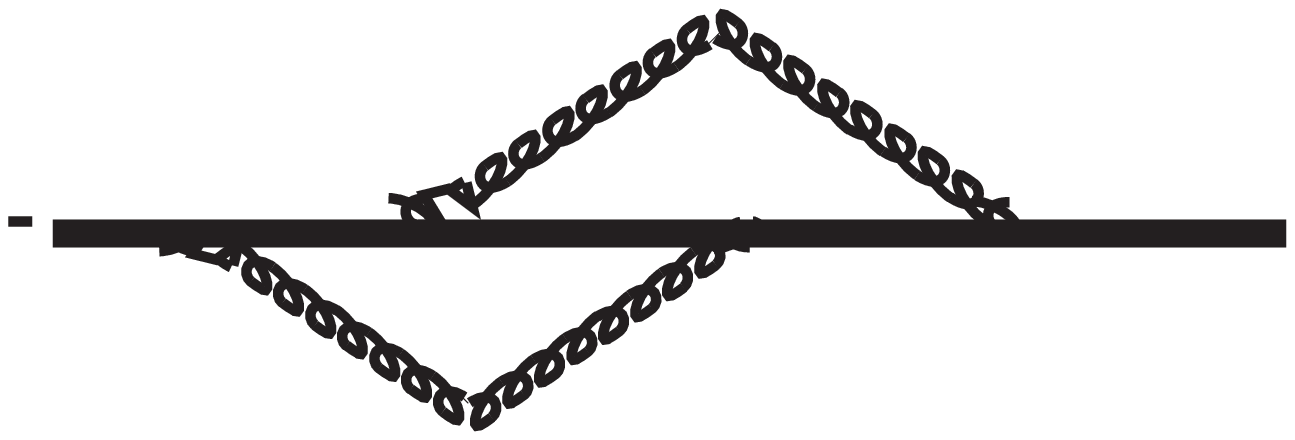}}\otimes\mbox{\bobfig{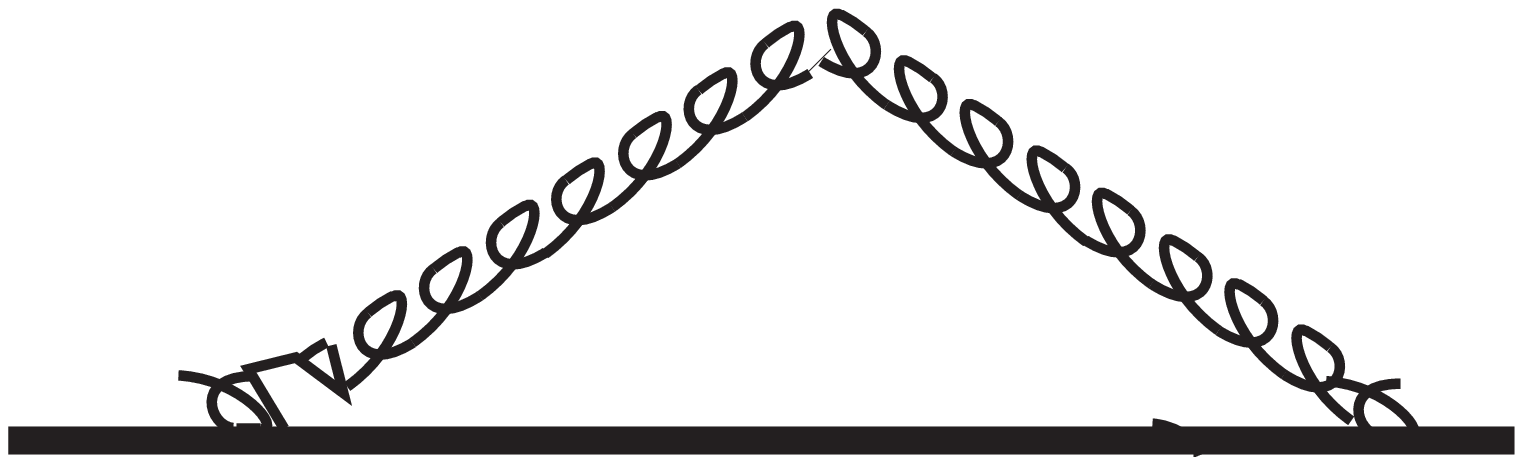}}
\end{eqnarray*}
The antipode $S_R[\bobfig{bob1.eps}]$
is then found to be
\begin{eqnarray*}
S_R[\bobfig{bob1.eps}] & = & -
R\left[\bobfig{bob1.eps}\right]
+
R\left[R\left[\bobfig{bob4.eps}\right]
\bobfig{bob5l.eps}\right]
+
R\left[R\left[\bobfig{bob4.eps}\right]
\bobfig{bob5r.eps}\right]\\
 & &
+
R\left[R\left[\bobfig{bob2.eps}\right]
\bobfig{bob3.eps}\right]
-
R\left[R\left[R\left[\bobfig{bob4.eps}\right]\bobfig{bob3.eps}\right]
\bobfig{bob3.eps}\right]
\end{eqnarray*}
Whenever we see $R$, this means we have to evaluate the analytic
expression corresponding to the Feynman graph on which $R$ acts
according to some chosen rules. For example, in minimal
subtraction we would introduce a regularization, say the
dimensional method, evaluate the bare Feynman graph as some
Laurent series in the corresponding regularization parameter, and
then read $R$ as the projection onto the pole term. On the other
hand, in on-shell BPHZ renormalization, we would introduce no
regularization, but let $R$ modify the integrand in a way such
that it fulfills certain constraints when external momenta are
on-shell. From the results in \cite{dkchen} we know that all such
schemes are fully equivalent. In particular the result of any
scheme can be recast into BPHZ form with appropriately chosen
constraints for external parameters.

Whatever scheme is chosen, the renormalized Green function
for the example finally becomes
\begin{eqnarray*}
{\bf \Gamma}_R[\bobfig{bob1.eps}] & = & (id-R)\left[\bobfig{bob1.eps}\right]
-
(id-R)\left[R\left[\bobfig{bob4.eps}\right]
\bobfig{bob5l.eps}\right]\\
 & & -
(id-R)\left[R\left[\bobfig{bob4.eps}\right]
\bobfig{bob5r.eps}\right]
-
(id-R)\left[R\left[\bobfig{bob2.eps}\right]
\bobfig{bob3.eps}\right]\\
 & &
+
(id-R)\left[R\left[R\left[\bobfig{bob4.eps}\right]\bobfig{bob3.eps}\right]
\bobfig{bob3.eps}\right]
\end{eqnarray*}
Here, $id$ is the identity map which leaves the bare diagram unchanged.
The same result is obtained when one works with rooted trees.

\section{Feynman diagrams as representations of rooted trees}
Particularly simple classes of Feynman graphs are given by graphs
which are represented by rooted trees which have the same
decoration at each vertex, and which give a multiplicative
representation defined below. We will soon see that the general
case can be reduced to similar representations, possibly allowing
for appropriate matrix structures to incorporate necessary
modifications due to spin. Hence it is legitimate to start the study
of the renormalization problem on graphs as simple as the ones shown
in Fig. \ref{f3}. These simplest graphs are completely governed by the
combinatorics of undecorated rooted trees.
Indeed, tree-factorials and vertex weights to be defined below are sufficient
to determine such graphs completely \cite{DiDa,dkchen}.

\subsection{Undecorated trees}
Let us first recapitulate some basic notions of rooted trees,
following \cite{dkchen}. For further details we refer the reader
to the Appendix. Let $T^{[0]}$ be the set of vertices of the
rooted tree $T$ with $n$ vertices. For any  vertex $v$ of $T$
which is not the root, let $t_v=P^c(T)$, where $c$ is the single
cut which only removes the one edge incoming to $v$ (all edges are
oriented away from the root). If $v$ is the root $r$, set $t_r=T$.
Also, $w(v)=\#(t_v)$ is the number of vertices of the tree $t_v$.
Then, we define the tree factorial by
\begin{equation}
T^!:= \prod_{v\in T^{[0]}}w(v).
\end{equation}
Hence, for the root, $w(r)=n=\#(T)$. Fig. \ref{f3} gives
instructive examples.

In the following, a variable $z$ plays the role of
a $q^2/\mu^2$ in field theory, and $x$ plays the role
of the regularization parameter, and can be identified
with $(4-D)/2$ in dimensional regularization.

We now  consider representations of ${\cal H}$ defined as follows.
Assume that we are given a set of functions $B_k(x)$ which are
Laurent series in the regularization parameter $x$, each having a
first-order pole. Using vertex weights as defined above we can
construct a function
\begin{equation}
\phi_z(t):=\prod_{v\in T^{[0]}} B_{w(v)}(x)z^{-nx}=B_t(x)z^{-nx},
\end{equation}
where $n$ is the number of vertices of $t$ and $z$ is to be
regarded as the scale parametrizing the representation. We have
also abbreviated $\prod_{v\in T^{[0]}} B_{w(v)}(x)=B_t(x)$. Quite
a number of interesting applications can be brought into this form
\cite{DiDa,dkchen}. Typically, one finds such representations
whenever one iterates a (scalar) Feynman diagram in terms of
itself, as described by a rooted tree \cite{DiDa,dkchen}. Even if
one only iterates one-loop integrals, such representations offer a
variety of interesting number-theoretic properties realized in the
same manner by different quantum field theories, ranging from the
absence of transcendentals in counterterms derived from ladder
graphs \cite{habil,bob1,bob2} (which represent trees of the form
$B_+^k(e)$, \cite{dkchen}) to the absence of non-rational
coefficients in quenched QED \cite{BDK} to similar phenomena when
summing up all diagrams with Connes Moscovici weights \cite{DiDa}.


Thus bare Green functions are given by
 $\phi_z(t)$.
Then we define the MS renormalization scheme by setting
\begin{equation}
R_{MS}[\phi_z]=<\phi_1>,
\end{equation}
where angle brackets denote projection onto the pole part of the
Laurent series in $x$ inside the brackets, and we set $z=1$ inside
these brackets. Next, let the minimal subtracted counterterm be
defined by
\begin{equation}
S_{R_{MS}}(\phi_z)(t)=-<\phi_z(t)
+\sum S_{R_{MS}}(\phi_z(t_{(1)})\phi_z(t_{(2)})>,
\end{equation}
which is the direct translation of (\ref{antip}) using the map
from Feynman graphs to decorated rooted trees, and let the
renormalized Green function be defined by
$\Gamma_{MS}(\phi_z)(t)=[S_{R_{MS}}\circ \phi_z\star \phi_z](t)$,
in accordance with (\ref{ren}), as expected. The reader should
have no difficulties confirming that for  $t_2$, the rooted tree
with two vertices,
\begin{equation}
S_{R_{MS}}(\phi_z)(t_2)=-<B_2B_1>+<<B_1>B_1>:=Z^{MS}_{t_2},
\end{equation}
and
\begin{equation}
\Gamma_{R_{MS},\phi_z}(t_2)=B_2 B_1z^{-2x}-<B_1>B_1z^{-x}-<B_2B_1>+<<B_1>B_1>.
\end{equation}
Therefore the antipode $Z_t^{MS}:=S_{R_{MS}}(\phi_z)(t)$ in MS reads,
for the first few trees,
\begin{eqnarray}
Z_{t_1}^{MS} & = & -<B_1>,\\ Z_{t_2}^{MS} & = &
-<B_2B_1>+<<B_1>B_1>,\\ Z_{t_{3_1}}^{MS} & = &
-<B_3B_2B_1>+<<B_1>B_2B_1>+<<B_2B_1>B_1>\nonumber\\
 & & -<<<B_1>B_1>B_1>,\\
Z_{t_{3_2}}^{MS} & = & -<B_3B_1^2>+2<<B_1>B_2B_1>\nonumber\\
 & & -<<<B_1>B_1>B_1>, \quad {\rm etc.}
\end{eqnarray}
which can be summarized by the formula \cite{dkchen}
\begin{equation}
Z_t^{MS}= \sum_{\mbox{\tiny full cuts $C$ of
$t$}}(-1)^{n_C}<\left[\prod_i <B_{t_i}>\right]B_{t_R}>.
\end{equation}

This being the result in the MS scheme, it was shown in
\cite{dkchen} how to express it in terms of a BPHZ type scheme,
where the use of a regularization scheme can be completely
avoided. Translated to the above representations, what we want to
achieve is the following: the above functions representing rooted
trees depend on a function $B_t(x)$, determined by a chosen tree,
a regularization parameter and the chosen one-loop graph (the
decoration) which is iterated according to the tree structure.
Further, there is the scale $z$, delivered in simple bare Feynman
diagrams typically by a variable, such as the square of a single
momentum, $q^2$. An on-shell scheme would simply subtract at a
chosen point $q^2/\mu^2=:z=1$, for example. An MS scheme would
also modify the Laurent series $B_t$, by projecting to $<B_t>$. In
the former case, one typically obtains the difference of
integrands evaluated at scales $z$ versus $1$, and hence one may
subtract integrands before having to regularize. In the latter
case, one modifies Laurent series after regularization and
integration. {\em We want to show how appropriate subtraction
points $\mu_t$ can be defined so that the MS renormalized
functions equal on-shell renormalized functions subtracted at
$\mu_t$}.

The idea is to use tree indexed parameters $\mu_t$. The antipode
$Z_t^\mu$ in a subtraction scheme using tree-indexed parameter sets
$\mu_t$ reads
\begin{equation}
Z_t^\mu= \sum_{\mbox{\tiny full cuts $C$ of $t$}}(-1)^{n_C}
\left[\prod_i B_{t^C_i} \mu_{t^C_i}^{-\#(t^C_i)
  x}\right]B_{t_R}\mu_{t_R}^{-\#(t_R) x},\label{ztmu}
\end{equation}
where $P^C(t)=\prod_i t_i$, which one immediately obtains by
standard formulas \cite{dkchen}. The first two trees
$t_1,t_2$ deliver immediately \begin{eqnarray} Z_{t_1}^\mu & = &
-B_{t_1}\mu_{t_1}^{-x}\\ Z_{t_2}^\mu & = &
-B_{t_2}\mu_{t_2}^{-2x}+B_{t_1}B_{t_1}\mu_{t_1}^{-2x},
\end{eqnarray}
as the reader should confirm using the equation (\ref{ztmu}).
For the definition of full cuts, appearing in the summation
above, we refer the reader to the Appendix.

Equating $Z_t^{MS}$ to $Z_t^\mu$ determines $\mu_t$ recursively
\cite{dkchen}:
\begin{equation}
\mu_t= \exp\left[\left(\frac{-1}{x t^!}\right)
\log(B^\prime_t/B_t)\right],
\label{mup}
\end{equation}
where
\begin{equation}
B^\prime:= \sum_{\mbox{\tiny full cuts $C$ of $t$}}(-1)^{n_C}
\left[\prod_i S_{R_{MS}}(\phi_z)(t^C_i)\right]S_{R_{MS}}(\phi_z)(t_R).
\end{equation}

One also confirms that the MS-renormalized Green function can now be
expressed through the on-shell subtracted one, where the on-shell condition
for $\phi_z(t)$ signifies subtraction at $z=\mu_t$ in this case. Thus
\begin{equation}
\Gamma_{R_{MS}}(\phi_z)(t)=\Gamma_{R_\mu}(\phi_z)(t)
\end{equation}
holds. Also, one immediately derives that $B^\prime_t/B_t=1+{\cal
O}(x)$, so that we properly arrive at well-defined conditions on the
external parameters $\mu_t$ to express the renormalized Green
function in MS through a BPHZ subtracted Green function with
tree-indexed boundaries.

\subsection{Decorated trees}
What is the situation that confronts us in general? In the
previous section, we dressed every vertex of a rooted tree with a
decoration $B_{w(v)}$, and then assigned the function
$\phi_z(t)=z^{-\#(t)x}\prod B_{w(v)}$ to a rooted tree, the
product being taken over all vertices of the tree. (Note that
$\#(t)=n=w(r)$, where $w(r)$ is the vertex-weight of the root.)
Clearly, the vertex weight $w(r)$ of a root of an undecorated tree
is simply the number $n$ of vertices of the tree. This equality does not
survive in the general case.

But if we only very slightly modify this equation and write
\begin{equation}
\phi_z(t)=z^{-w(r) x}\prod_v B_{w(v)}.\label{master}
\end{equation}
we obtain the form which permits all necessary generalizations.
Such generalizations include the following cases:
\begin{itemize}
\item In general, decorations can be provided by analytic expressions of any loop order.
We will have to attach appropriate weights to the vertices
themselves, and hence will have to generalize the concept of
vertex weights.

The crucial step is to devise some sensible notation.
An arbitrary rooted tree has vertices decorated by
graphs corresponding to analytic expressions without
subdivergences which themselves can have any loop order. We still
assume that these analytic expressions can be described by scalar
functions. Loop integrations of decorations at vertices below the
vertex considered will alter the measure by an amount
$(k^2)^{w(v_i)}$ where $v_i$ is the generalized vertex weight
defined in Fig. \ref{gvw} and $k$ is an appropriate external
momentum for the decoration at $v_i$.
At each vertex we indicate the loop number of its decoration.
The generalized vertex weight is then the sum of all these loop
numbers, assigned to the vertex in question and all vertices below it.
>From this, one obtains the generalized tree factorial, which is to be
used in (\ref{master}).

\item We might consider graphs which do not come from scalar theories.
In order to cope with form factors, one needs to devise an
appropriate matrix calculus; this will be worked out in a detailed
example in a section below. Essentially, the master equation
(\ref{master}) remains valid, but becomes now a matrix equation.
This is the most non-trivial generalization, and we will discuss
it in detail presently.

In the same manner, one can use matrices to keep track of internal
structure of a graph:  at some stage one has to bookkeep the
location of subdivergences in a larger graph, and this leads to a
matrix calculus which is a straightforward generalization of the
example which we consider below. Indeed, one can enumerate all
vertices of a graph and can construct suitable matrices which
store the information about the nature of these vertices and the
propagators connecting them. In this manner, the previously
discussed changes of measures due to subdivergences operate on
entries of such matrices, and it is a straightforward notational
exercise to set-up a convenient calculus. We will not consider the
most general case here, but rather present below an example for
QED which should make the idea sufficiently clear.

\item In general, divergent subgraphs depend on masses and more than
one external momentum; hence on multiple scales. For example, a
vertex subdivergence will depend on more than one internal
momentum usually and this can spoil the multiplicative form of
(\ref{master}). This difficulty can be solved by a proper
decomposition into sums of expressions, each of the form
(\ref{master}). This was described already in \cite{hopf}. One
proceeds by decomposing any function of multiple scales into one
of a single scale and another finite multiple scale function. This
is always possible by adding zero in appropriate ways.
\end{itemize}

The final result is that from a local QFT we essentially get analytic
expressions which can be expressed in the form (\ref{master}).
We now utilize this fact to show that they essentially behave as
generalized iterated integrals.

\section{Feynman diagrams from generalized iterated integrals}
Now that we have established that (\ref{master}) is an acceptable
prototype expression into which Feynman diagrams can be resolved,
we want to utilize the fact that the simplest representation of
such form can be obtained from generalized iterated integrals.
Such generalized iterated integrals were prominent in
\cite{dkchen} and give a convenient means to investigate
renormalization. Generalized iterated integrals are distinguished
from ordinary iterated integrals by the fact that boundaries
of the domains of integration are determined as a function
of the rooted tree which the integral represents.
In this respect,
ordinary iterated integrals use a function which is constant
on rooted trees. For such a generalization,
the shuffle symmetry of iterated integrals
is lost.

The extra freedom gained by using this generalization
will allow us to rewrite Feynman diagrams as such generalized
iterated integrals. We will see that the generalization only affects
non-leading UV divergences, and hence we will obtain later the result
that the leading UV divergences of Feynman integrals still obey a
shuffle identity.

At this point, it might be worth stressing that if we were able
to reduce Feynman diagrams completely to ordinary iterated integrals
we would obtain considerable progress in understanding the analytical
and transcendental structure of Green functions, as the theory
of ordinary iterated integrals is quite rich and highly developped.
We hope that a similar theory will be emerging in the future
for generalized iterated integrals.
\subsection{Iterated Integrals}
Essentially, we want to go one step further as in \cite{dkchen}
and express contributions to Green functions in QFT as iterated
integrals. We will start our considerations by reminding ourselves
of some basic properties of iterated integrals \cite{chen,ShSt},
specializing to the case of a single function $f(x)$ with
associated one-form $f(y)dy$ on the real line. We assume that
$f(y)$ behaves $\sim 1/y$ for large $y$.

Iterated integrals built with the help of $f$ are parametrized by an
integer $n$ and two real numbers $a,b$ say. They are defined by
\begin{eqnarray}
F^{[0]}_{a,b} & = & 1,\qquad \forall a,b\in {\bf R},\label{triv}\\
F^{[n]}_{a,b} & = & \int_a^b f(x)F^{[n-1]}_{a,x}dx,\qquad\forall
n>0.
\end{eqnarray}
Hence we can write them as an integral over the simplex,
\begin{equation}
F_{a,b}^{[n]}=\int_{a\leq x_1<\ldots<x_{n}\leq b}
f(x_1)\ldots f(x_n) dx_1\ldots dx_n.
\end{equation}
We will also consider \[ F^{[T]}_{a,b}=\int_a^b dx f(x) \Pi_i
F^{[t_i]}_{a,x}
\]
where $T$ is a rooted tree with $B_-(T)=\Pi_i t_i$.

A generalization which turns out to be quite useful in practice is
to let even the  boundaries  $a,b$ be indexed by decorated rooted
trees. This is similar to what we did before when we expressed the
results of an MS calculation by an on-shell scheme with suitably
adjusted tree-dependent parameters. Here, parameters are provided
by boundaries in the iterated integral.  Hence, we define an
iterated integral which serves as a bare Green function
\begin{equation}
G_{b_t,\infty}(t)=\int_{b_t}^\infty f(x)\prod G_{x,\infty}(t_i)dx.
\end{equation}
The product is again over all branches $t_i$ of $B_-(t)$, and we
set $G_{b_e,\infty}(e)=1$. This is the same as before, except that
we now label the lower outer boundary by a decorated rooted tree
$t$ and set the upper boundary to infinity. We assume that a
renormalization map $R_a$ shifts the outer lower boundary $b_t$ to
another value $a_t$, and that the coproduct action extends to this
label. We define $\phi_b$ to be the map which sends $t\to
G_{b_t,\infty}(t)$.

It is worthwile citing here an example taken from \cite{dkchen}
which exhibits the intricate nature of a change of scales in full
generality. For the rooted tree with two vertices, $t_2$, having coproduct
\begin{equation}
\Delta(t_2)=t_2 \otimes e+e\otimes t_2+t_1\otimes t_1
\end{equation}
we formally obtain a counterterm
$$S_{R_a}(\phi_b)(t_2)=-R_a[\phi_b(t_2)+m[(S_{R_a}\circ\phi_b\otimes
\phi_b)\Delta^\prime(t_2)]]$$ (where
$\Delta^\prime(T)=\Delta(T)-[1\otimes T]-[T\otimes 1]$) as
\begin{equation}
S_{R_a}(\phi_{b})(t_2)= \left[ \int_{a_{t_1}}^\infty
\int_{a_{t_1}}^\infty-\int_{a_{t_2}}^\infty\int_x^\infty\right]f(x)f(y)dydx.
\end{equation}
and a renormalized iterated integral
$\Gamma_{a,b}(t_2)=m[(S_{R_a}\circ\phi_b\otimes
\phi_b)\Delta(t_2)]$ as
\begin{equation}
\Gamma_{a,b}(t_2)=\left[\int_{b_{t_2}}^\infty \int_x^\infty-
\int_{b_{t_1}}^\infty \int_{a_{t_1}}^\infty+ \int_{a_{t_1}}^\infty
\int_{a_{t_1}}^\infty-\int_{a_{t_2}}^\infty\int_x^\infty\right]f(x)f(y)dydx.
\end{equation}
In this notation, $a,b$ are to be regarded as representing actually a whole
set of constants $a_t,b_t$, parametrizing the relevant scales
for the decorated tree under consideration.

We now want to change the set of scales and reexpress how
$\Gamma_{a,b}$ can be obtained from renormalized Green functions
$\Gamma_{a,s}$ and $\Gamma_{s,b}$. Obviously, the convolution
\[
\Gamma_{a,b}(t)=[\Gamma_{a,s}\star\Gamma_{s,b}](t)
\]
holds. This form of Chen's Lemma describes what happens if we
change the renormalization point. It works for the simple case of
iterated integrals in the same manner as for full Green functions
of QFT. $\Gamma_{a,b}=\Gamma_{a,s}\star\Gamma_{s,b}$ now becomes
\begin{eqnarray*}
 & & \left[
\int_{b_{t_2}}^\infty \int_x^\infty- \int_{b_{t_1}}^\infty
\int_{a_{t_1}}^\infty+ \int_{a_{t_1}}^\infty
\int_{a_{t_1}}^\infty-\int_{a_{t_2}}^\infty
\int_x^\infty\right]f(x)f(y)dydx\\
 & = &
\left[ \int_{s_{t_2}}^\infty \int_x^\infty- \int_{s_{t_1}}^\infty
\int_{a_{t_1}}^\infty+ \int_{a_{t_1}}^\infty
\int_{a_{t_1}}^\infty-\int_{a_{t_2}}^\infty\int_x^\infty\right]
f(x)f(y)dydx\\
 &  &
+\left[\int_{b_{t_2}}^\infty \int_x^\infty- \int_{b_{t_1}}^\infty
\int_{s_{t_1}}^\infty+ \int_{s_{t_1}}^\infty \int_{s_{t_1}}^\infty
-\int_{s_{t_2}}^\infty\int_x^\infty\right]f(x)f(y)dydx\\
 & &
+\left[\int_{s_{t_1}}^\infty -\int_{a_{t_1}}^\infty\right]f(y)dy
\left[\int_{b_{t_1}}^\infty -\int_{s_{t_1}}^\infty\right] f(x)dx.
\end{eqnarray*}
This is evidently true, as the reader may readily check.

Finiteness of $\Gamma_{a,b}$ now imposes conditions on the
tree-indexed parameters, a fact which we  will utilize in the next
section. Actually, we will determine these parameters in a
way which reproduces the results of a renormalized QFT.

\section{Counterterms and renormalized Green functions as iterated
integrals}
We are now in the fortunate position of being able to transform Feynman
diagrams to generalized iterated integrals. To that end,
let us consider the simplest case of an iterated integral
based on the single one-form  $f(y)dy=y^{-1-x}dy$,
which diverges logarithmically when integrated to infinity
and $x\to 0$.

Then, we obtain the bare iterated integral,
\[ G_{b,\infty}(t)=\frac{1}{t^!x^{-\#(t)}}b_t^{-\#(t)x}, \]
and associated counterterms $S_{R_\mu}(\phi_b)(t)$,
which in this case are simply
\[ S_{R_\mu}(t)= -R_\mu[\phi_b(t)+m[(S_{R_\mu}\otimes
id)(\phi_b\otimes \phi_b))\Delta^\prime]]=\phi_\mu(S(t)),\]
with $R_\mu(\phi_b)=\phi_\mu$. Thus we arrive at the renormalized integral
\[
\Gamma_{\mu,b}(t)= \phi_\mu(P^C(t))\phi_b(R^C(t)).\]
Explicitly,
\[
S_{R_\mu}(t_2)=-\frac{1}{2x^2}\mu_{t_2}^{-2x}+\left[\frac{1}{x}\mu_{t_1}^{-x}\right]^2
\]
and
\[
\Gamma_{\mu,b}(t_2)=\frac{1}{2x^2}b_{t_2}^{-2x}-\frac{1}{x^2}b_{t_1}^{-x}\mu_{t_1}^{-x}
+\frac{1}{x^2}\mu_{t_1}^{-x}\mu_{t_1}^{-x}-\frac{1}{2x^2}\mu_{t_2}^{-2x},
\]
which is nonsingular when $x\to 0$.

Next determine the set of parameters $\mu_t$ by iteratively setting
\[
S_{R_\mu}(t)=S_{R}(\phi_z(t)),
\]
wherein we identify the parameter $x$ in
the iterated integrals with the regularization parameter for bare
Green functions in QFT and use the representation (\ref{master})
for those Green functions. A similar equation is used to determine
the set of parameters $b_t$ in the iterated integrals such that
\[
\Gamma_{\mu,b}(t)={\bf \Gamma}_R(\phi_z(t)).
\]
If we use a on-shell
subtraction at the point $z=1$ for example, one simply finds
\[
\mu_t=\exp\left(
\frac{1}{[\#(t)x]}\log[t^!x^{\#(t)}B_t(x)]\right).
\] Note that this quantity has no essential singularity at
$x=0$ due to the fact that $t^!x^{\#(t)}B_t$ has the form $1+{\cal
O}(x)$ in our adopted normalization.
In this simple on-shell scheme, $b_t$ turns out to be
\[ b_t=\mu_t z^{-\#(t)x} \]
and the resulting generalized iterated integral is a well-defined
finite quantity which reproduces the renormalized Green function.

Changing  renormalization conditions or schemes is now evidently a
convolution of iterated integrals. For example, transforming to a
MS scheme amounts to a convolution similar to the one described
before, and we would eventually have
\[\Gamma_{\mu^{MS},b}(t)=[\Gamma_{\mu^{MS},\mu}\star
\Gamma_{\mu,b}](t).\]
The second Green function on the rhs uses
the same $\mu_t,b_t$ as defined above, the first on on the rhs
uses the same $\mu_t$ but determines $\mu^{MS}_t$ using minimal
subtracted QFT Green functions. Such convolutions were already
applied in \cite{DiDa} and systematically derived in
\cite{dkchen}. They essentially reduce the calculation of
renormalized Green functions to a determination of appropriate
functions $B_t$ from which all other results follow via the
algebraic structure of the Hopf algebra. Note that the convolution
also permits us to describe the change of scales in bare Green
functions; for example $\phi_z(t)\to \phi_{\rho z}(t)$ is described by
\[ \phi_{\rho z}(t)=[\phi_z\star \Gamma_{R_z}(\phi_{\rho z})](t), \]
which immediately translates to iterated integrals as the convolution
$\Gamma_{\mu,b^\prime}=\Gamma_{\mu,b}\star\Gamma_{b,b^\prime}$
where $b^\prime_t=b_t\rho^{-\#(t)x}$.

In the following we determine some of the $B_t$ explicitly, to illustrate
the methods.

\subsection{Massless Yukawa theory}
Let us apply the previous ideas to massless Yukawa theory, with
interaction
\[g_0\bar{\psi}_0\gamma_5\psi_0\phi_0 =
Z_1g\bar{\psi}\gamma_5\psi\phi.\] For illustrative purposes we
consider a three-loop contribution to the fermion self-energy and
an analogous one for the vertex part, as sketched in Fig. \ref{rain}.

In order to describe the results, in the context of dimensional
regularization, the parameter $x$ is identified with
$(4-D)/2\equiv \epsilon$ and $z$ with the scale factor
$\mu^2/p^2$, where $p$ is the external momentum and $\mu$ is the
renormalization mass scale which arises via the dimensionality of
the bare coupling: $g_0^2\equiv(-4\pi)^{-x}\mu^{2x}
4\pi\alpha/\Gamma(1-x)$. The calculations bring in supplementary
integrals which are readily calculated in any dimension $D=4-2x$,
namely
\begin{mathletters}
\begin{equation}
 g_0^2I(p,nx)\equiv -ig_0^2\int \frac{d^Dk/(2\pi)^D}{k^2((k+p)^{2(1+nx)}}
  = \frac{\alpha z^x}{4\pi} \frac{i_{n+1}}{(p^2)^{nx}},
\end{equation}
where
\begin{equation}
i_{n+1}\equiv \frac{\Gamma(1-(n+1)x)\Gamma((n+1)x)}
                   {\Gamma(1+nx)\Gamma(2-(n+2)x)},
\end{equation}
\end{mathletters}
and
\begin{mathletters}
\begin{equation}
 g_0^2J(p,nx)\equiv -ig_0^2\int \frac{[1+p.k/p^2]d^Dk/(2\pi)^D}
  {k^2((k+p)^{2(1+nx)}}
   = \frac{\alpha z^x}{4\pi} \frac{j_{n+1}}{(p^2)^{nx}},
\end{equation}
where
\begin{equation}
j_{n+1}\equiv \frac{\Gamma(2-(n+1)x)\Gamma((n+1)x)}
                   {\Gamma(1+nx)\Gamma(3-(n+2)x)}.
\end{equation}
\end{mathletters}
Those functions $i_{n+1},j_{n+1}$ play the role of the functions
$B_{w(v)}$ of the previous considerations. Applying the on-shell
subtraction which sets $z\to 1$, $R_{1}$,
to this graph, yields the following result for the
self-energy integral:
$$\left(\frac{\alpha}{4\pi}\right)^3\left[(z^{3x}-1)j_1^2j_3 -
   2(z^{2x}-1)j_1^2j_2 + (z^x-1)j_1^3\right];\qquad z\equiv \mu^2/p^2,$$
which is obtained as  $m[(S_{R_1}\otimes id)\Delta(j_t(z))]$
   where $j_t(z)=\prod_v j_{w(v)}[\alpha z^{-x}]^{\#(t)}$
plays the role of $\phi_z(t)$; $\Delta$ acts on the tree parametrizing
$j_t$ such that $\Delta(j_t(z))=\sum j_{t_{(1)}}(z)\otimes j_{t_{(2)}}(z)$
and $R_1[j_t(z)]=j_t(1)$.

Similarly, the following result for the vertex integral at zero
meson momentum is obtained:
$$-\left(\frac{\alpha}{4\pi}\right)^3\left[(z^{3x}-1)j_1^2i_3 -
   2(z^{2x}-1)j_1^2i_2 + (z^x-1)j_1^2i_1\right];\qquad z\equiv \mu^2/p^2,$$
   in which we evidently see, as before, the $(id-R)$ structure
   of renormalized Green functions.
The limit $x\rightarrow 0$ may readily be taken in the renormalized
answers, producing
$$ \left(\frac{\alpha}{4\pi}\right)^3\left[\frac{U^3}{24} + \frac{U^2}{16}
                   + \frac{U}{16}\right];\qquad U\equiv \ln(\mu^2/p^2) $$
for the self-energy and
$$ -3\left(\frac{\alpha}{4\pi}\right)^3\left[\frac{U^3}{12}+\frac{U^2}{4}
                   + \frac{U}{2}\right]; \qquad U\equiv \ln(\mu^2/p^2) $$
for the vertex part. The latter carries an extra weight factor 3
because of the way that the interior self-energies can be
distributed.

\subsection{Quantum electrodynamics}
In this case the renormalization scale $\mu$ enters through
$e_0^2\equiv(-4\pi)^{-x}\mu^{2x}4\pi\alpha/\Gamma(1-x)$, where $\alpha$ now
has the significance of the fine structure constant. Because we are dealing
with the massless version, the self-energy assumes the form, $\Sigma(p)
= \not{p} A(p^2)$, while the vertex at zero-momentum transfer contains
two terms (form-factors):
$$\Gamma_\mu(p,p) = \gamma_\mu{\cal F}(p^2) -
 \not{p}\gamma_\mu\!\not{p}{\cal G}(p^2)/p^2,$$
of which only the first requires (infinite) renormalization. The form
factors can be found in the usual manner by tracing appropriately:
\begin{eqnarray}
{\rm Tr}[\gamma\cdot\Gamma]&=&2^{2-x}[(4-2x){\cal F}-x{\cal G}]\nonumber\\
{\rm Tr}[\not{p}p\cdot\Gamma]&=&p^22^{2-x}[{\cal F} - {\cal G}],\nonumber
\end{eqnarray}
and it is helpful to define the auxiliary integral
$$e_0^2H(p,nx)=\frac{\alpha(1-x)z^xh_{n+1}}{4\pi(p^2)^{nx}},$$
where
$$h_{n+1}\equiv -\frac{(1-x)\Gamma((n+1)x)\Gamma(1-(n+1)x)}
                       {2\Gamma(2+nx)\Gamma(3-(n+2)x)}.$$
We should also exhibit
a combination which features prominently in actual
computations, viz.
$$i_{n+1} + Dh_{n+1} = \Gamma(2-(n+1)x)\Gamma(1+(n+1)x)/
                            \Gamma(3-(n+2)x)\Gamma(2+nx);\quad D=4-2x.$$

Lowest order in $\alpha$ calculations give ($z\equiv \mu^2/p^2$)
\begin{mathletters}
\begin{equation}
 A = \left(\frac{\alpha}{2\pi}\right)j_1 z^x,
\end{equation}
\begin{eqnarray}
 {\cal F} &=& (1-x)^2 \left(\frac{\alpha}{2\pi}\right)h_1 z^x,\nonumber\\
 {\cal G} &=& \left(\frac{\alpha}{2\pi}\right)[i_1+Dh_1] z^x,
\end{eqnarray}
\end{mathletters}
leading simply to the on-shell subtracted expressions,
\begin{mathletters}
\begin{equation}
 A_{os} = \left(\frac{\alpha}{2\pi}\right)j_1 [z^x-1],
\end{equation}
\begin{equation}
 \Gamma_{\mu,os} = \left(\frac{\alpha}{2\pi}\right)j_1\left[
                   \gamma_\mu(z^x-1) - 2x\frac{p_\mu}{\not{p}}z^x\right].
\end{equation}
\end{mathletters}

It is very easy to continue this process to higher orders for the
self-energy function $A(p^2)$ in the Feynman gauge say; all we
need do is to substitute $g^2$ in massless Yukawa theory by $2e^2$
in QED, the factor of $2$ arising from the gamma-matrix trace.
However the vertex computations are subtler, because they involve
the two form factors ${\cal F}$ and ${\cal G}$. Here we require
knowledge of the generic integral,
\[ie_0^2\int\frac{d^D\!K}{(4\pi)^Dk^2}\gamma_\nu\frac{1}{\not{p}+\not{k}}
 \left[ f_n\gamma_\mu - g_n\frac{(\not{p}+\not{k})\gamma_\mu(\not{p}+\not{k})}
 {(p+k)^2}\right]\left(\frac{\mu^2}{(p+k)^2}\right)^{nx}
 \frac{1}{\not{p}+\not{k}}\gamma^\nu\]
\begin{equation}
 \equiv\left(\frac{\alpha}{2\pi}\right)\left[f_{n+1}\gamma_\nu - g_{n+1}
 \frac{(\not{p}+\not{k})\gamma_\mu(\not{p}+\not{k})}{(p+k)^2}\right]
 \left(\frac{\mu^2}{p^2}\right)^{(n+1)x},
\end{equation}
starting with $f_0=1,g_0=0$. In this way we straightforwardly arrive at the
recurrence property,
\[
{\bf v}_{n+1}={\bf M}_{n+1}{\bf v}_n,
\]
where ${\bf v}_i$ is a column vector $(f_{i},g_{i})^T$ with ${\bf
v}_0=(1,0)^T$ and ${\bf M}_n$ is a $2\times 2$ matrix with entries
$(M_n)_{rs}$ given by
\begin{eqnarray}
 (M_n)_{11}\equiv \frac{(1-x)^2\Gamma(nx)\Gamma(1+nx)}
                                {\Gamma(2+(n-1)x)\Gamma(3-(n+1)x)}, & &
                \,  (M_n)_{12}\equiv \frac{\Gamma(1-nx)\Gamma(nx)}
                                {\Gamma(1+(n-1)x)\Gamma(2-(n+1)x)} \\
                \,  (M_n)_{21}\equiv \frac{\Gamma(2-nx)\Gamma(1+nx)}
                                {\Gamma(2+(n-1)x)\Gamma(3-(n+1)x)}, & &
                \,  (M_n)_{22} = 0.
\end{eqnarray}

We now introduce the matrix-function ${\bf R}$ with entries
$R_{11}=R_1$ and 0 elsewhere, where $R_1$ is again the map which sends
$z=\mu^2/p^2 \to 1$. Further, we use the $2\times 2$ unit matrix ${\bf I}$
such that an insertion of a fermion self-energy amounts to an insertion of
matrices of the type ${\bf A}_k:={\bf I}\frac{\alpha}{2\pi}j_k$ into
a string of matrices. Then we can express the result for any
diagram $\Gamma(t)$, which iterates one-loop vertices at
zero-momentum transfer and also has various combinations of
one-loop fermion self-energies as subdivergences, by the formula
\[ \Gamma(t)=\left[\prod_{v\in t^{[0]}} B_{w(v)}^{i(v)}z^{-w(r)x}\right], \]
where $t$ is a decorated rooted tree allowing for one of two
possible decorations $i(v)$ at each vertex: either $i(v)$
evaluates to a one-loop vertex function or to a one-loop fermion
self-energy. As both decorations are of one-loop order, $w(v)$ is
the usual vertex weight discussed before. We then set
$B_{j}^{i(v)}:= {\bf M}_j$ if the vertex $v$ of $t$ is decorated
by the one-loop vertex-function, and $B_j^{i(v)}={\bf A}_j$ if the
vertex $v$ is decorated by a one-loop fermion self-energy.
Various examples are drawn in Fig. \ref{matrix}.

Renormalization now amounts to inserting the matrix ${\bf R}$ into the
string of matrices at places which correspond to cuts at the
decorated tree. Hence the formulae for the counterterms and
renormalized Green functions remain essentially unchanged.
For the Feynman graph with the tree $t$ shown in Fig. \ref{matrix},
we find the bare Green function
\[
\Gamma(t)={\bf A}_1{\bf A}_1 {\bf M}_3 {\bf v}_0z^{-3x}.
\]
The counterterm being $Z_R=-{\bf R}[\Gamma(t)+m((Z_R\otimes id)
(\Gamma\otimes\Gamma)\Delta^\prime(t))]$, we get
\[ -{\bf R}[{\bf A}_1{\bf A}_1 {\bf M}_3 {\bf v}_0]+2 {\bf R}[{\bf
A}_1]{\bf R}[{\bf A}_1{\bf M}_2{\bf v}_0]-{\bf R}[{\bf A}_1] {\bf
R}[{\bf A}_1] {\bf R}[{\bf M}_1 {\bf v}_0]. \]
Finally the renormalized Green function $m[(Z_R\otimes
id)(\Gamma\otimes \Gamma)\Delta(t)]$ leads to the on-shell
subtracted result (aside from a weight factor of 3)
$$\left(\frac{\alpha}{2\pi}\right)^3 j_1^2\left[\gamma_\mu\left(
(M_3)_{11}(z^{3x}-1)-2(M_2)_{11}(z^{2x}-1)+(M_1)_{11}(z^x-1)\right.\right.$$
$$\left.\left. \qquad -  (M_3)_{12}+2(M_2)_{12}-(M_1)_{12}\right)\right. $$
$$\left.\qquad\qquad-\left((M_3)_{12}z^{3x}-2(M_2)_{12}z^{2x}+(M_1)_{12}
z^x\right) \not{p}\gamma_\mu\!\not{p}/p^2\right].$$
The renormalized contribution may be found by proceeding to the delicate limit
$x\rightarrow 0$, when the above answer collapses into
$$-\left(\frac{\alpha}{2\pi}\right)^3\left[\gamma_\mu
   \frac{2U^3+3U^2+15U-9}{48} -\frac{\not{p}\gamma_\mu\!\not{p}}{p^2}
   \frac{2U^2-2U+3}{16}\right];\quad U=\ln(\mu^2/p^2).$$

Other diagrammatic contributions may be extracted in much the same
way; we have evaluated many more results for the renormalized
self-energy $A_{os}$ and the vertex at zero momentum transfer
$\Gamma_{\mu,os}$. A useful check on the calculations, that we
have carried out, is to verify the Ward identity,
$\Gamma(p,p)=\partial\Sigma(p)/\partial p_\mu$, since a particular
contribution to $A$ will produce a combination of vertex graphs
with appropriate weights.

\subsection{A relaxed shuffle product}
Proper iterated integrals satisfy a shuffle identity \cite{chen,ShSt}.
Due to this shuffle identity, one can express all tree-iterated integrals
representing rooted trees as proper iterated integrals. The latter
are based on trees without sidebranching and form a
closed Hopf subalgebra of ${\cal H}$.
The generalized iterated integrals which we used to summarize
some restricted results of QFT do not obey a shuffle algebra, and hence we get
non-trivial representations of the full Hopf algebra \cite{dkchen}.

However one can easily establish that the leading coefficient of
bare Feynman diagrams still obeys such an identity,
by making use of the expression in terms of generalized iterated
integrals. For that it suffices to realize that if the $\mu_t$ become
$t$-independent, we get a proper iterated integral. But to leading order,
we have
\begin{equation}
B_t=\frac{1}{t^!x^{\#(t)}}F(t),
\end{equation}
with $F(t)=1+{\cal O}(x)$, by definition; so to leading order,
$F(t)=1$ $\forall t$, and then $B_t=1/t^!x^{\#(t)}$, which
can be easily obtained as the iterated integral $G_{1,\infty}(t)$
defined previously. This explains relations known to practitioners
of QFT that hold between leading coefficients of the overall
divergence of graphs representing different trees.

\section{Comments and Conclusions}
In this and previous papers we have developed a number of mathematical
tools which clarify and simplify the renormalization procedure for any
renormalizable quantum field theory. These are encapsulated in formulae
such as  (11), (13), (15), (26) and (30); we have provided two nontrivial
examples where these ideas can be fruitfully applied without too much
effort. Therefore, it is our firm belief and hope that all these tools
will enable the practitioner of QFT to obtain the amplitudes associated
with renormalized Feynman diagrams by purely combinatorial means.
We also anticipate that the connection between these amplitudes
and iterated one-dimensional integrals, whose endpoints define the
renormalization scheme, will also lead to substantial progress in
automating the method, since they provide a significant step along this
direction. The mathematical tools emphasize the fundamental connection
between geometry, topology, number theory and Feynman diagrams that is
emerging \cite{DD,Br,habil,pisa,BGK,BK15}.

\acknowledgements Thanks are due to David Broadhurst and Alain
Connes for interest and discussions. D.K.~thanks the
I.H.E.S.~(Bures-sur-Yvette) for friendly hospitality and marvelous
working conditions and thanks the University of Tasmania for
hospitality during a visit March-April 1998. D.K.also thanks the
DFG for support by a Heisenberg Fellowship.

\section*{Appendix}
\subsection{The Hopf Algebra}

We follow \cite{CK,overl} closely. A {\em rooted tree} $t$ is a
connected and simply-connected set of oriented edges and vertices
such that there is precisely one distinguished vertex which has no
incoming edge. This vertex is called the root of $t$. Further,
every edge connects two vertices and the {\em fertility} $f(v)$ of
a vertex $v$ is the number of edges outgoing from $v$. The trees
being simply-connected, each vertex apart from the root has a
single incoming edge.

As in \cite{CK}, we consider the (commutative) algebra of
polynomials over ${\bf Q}$ in rooted trees; hence the
multiplication $m(t,t^\prime)$ of two rooted trees means drawing
them next to each other in arbitrary order.
Observe that any rooted tree $t$ with root $r$ yields $f(r)$
trees $t_1$, $\ldots$, $t_{f(r)}$ which are the trees attached to
$r$. The unit element of this algebra is $1$, corresponding, as a
rooted tree, to the empty set.

Let $B_-$ be the operator which removes the root $r$ from a tree
$t$:
\begin{equation}
B_-: t\to B_-(t)=t_1 t_2\ldots t_{f(r)}.
\end{equation}
Fig. \ref{guill-} depicts an example. Also let $B_+$ be the
operation which maps a monomial of $n$ rooted trees
to a new rooted tree $t$ which has a root $r$ with fertility
$f(r)=n$ which connects to the $n$ roots of $t_1,\ldots,t_n$.
\begin{equation}
B_+: t_1\ldots t_n\to B_+(t_1\ldots t_n)=t.
\end{equation}
This is clearly the inverse to the action of $B_-$.
For any rooted tree $t$ one has
\begin{equation}
B_+(B_-(t))=B_-(B_+(t))=t,
\end{equation}
and Fig. \ref{guill+} provides one such example.
We further set $B_-(t_1)=1$, $B_+(1)=t_1$.

We will introduce a Hopf algebra on such rooted trees by taking the
opportunity to cut such trees in pieces. We start with the most
elementary possibility. An {\em elementary cut} is a cut of a
rooted tree at a single chosen edge, as indicated in
Fig. \ref{ecut}.

Before introducing the coproduct we  finally introduce the notion
of an {\em admissible cut}, also called a {\em simple cut}. It is
any assignment of elementary cuts to a rooted tree $t$ such that
any path from any vertex of the tree to the root has at most one
elementary cut. Fig. \ref{cut} depicts such a situation.
An admissible cut $C$ maps a tree to a monomial in trees. If the
cut $C$ contains $n$ elementary cuts, it induces a map
\begin{equation}
C: t\to C(t)=\prod_{i=1}^{n+1} t_{j_i}.
\end{equation}

Note that precisely one of these trees $t_{j_i}$ will contain the
root of $t$. Let us denote this distinguished tree by $R^C(t)$.
The monomial which is delivered by the $n-1$ other factors is
denoted by $P^C(t)$.
The definitions of $C,P,R$ can be extended to monomials of trees
in the obvious manner, by choosing a cut $C^i$ for every tree
$t_{j_i}$ in the monomial:
\begin{eqnarray*}
C(t_{j_1}\ldots t_{j_n}) & := & C^1(t_{j_1})\ldots C^n(t_{j_n}),\\
P^C(t_{j_1}\ldots t_{j_n}) & := & P^{C^1}(t_{j_1})\ldots
P^{C^n}(t_{j_n}),\\ R^C(t_{j_1}\ldots t_{j_n}) & := &
R^{C^1}(t_{j_1})\ldots R^{C^n}(t_{j_n}).
\end{eqnarray*}

Let us now establish the Hopf algebra structure. Following
\cite{hopf,CK} we first define the counit and the coproduct. The
{\em counit} $\bar{e}$: ${\cal A} \to {\bf Q}$ is simple:
$$\bar{e}(X)=0 $$ for any $X\not= 1$, $$ \bar{e}(1)=1. $$

The {\em coproduct} $\Delta$ is defined by the equations
\begin{eqnarray}
\Delta(1) & = & 1\otimes 1\\ \Delta(t_1\ldots t_n) & = &
\Delta(t_1)\ldots \Delta(t_n)\\ \Delta(t) & = & t \otimes 1
+(id\otimes B_+)[\Delta(B_-(t))],\label{cop2}
\end{eqnarray}
which defines the coproduct on trees with $n$ vertices iteratively
through the coproduct on trees with a lesser number of vertices.
See Fig. 11 for an example. Actually, the coproduct can be written
as \cite{hopf,CK}
\begin{equation}
\Delta(t)=1\otimes t+ t\otimes 1+ \sum_{\mbox{\tiny adm.~cuts $C$
of $t$}}P^{C}(t)\otimes R^C(t)=:1\otimes t+t\otimes
1+\Delta^\prime(t),\label{cop1}
\end{equation}
which defines $\Delta^\prime$.

Up to this point we have established a bialgebra structure, but it is
actually a Hopf algebra. Following \cite{hopf,CK} we find the
antipode $S$ as
\begin{eqnarray}
S(1) & = & 1\\ S(t) & = & -t-\sum_{\mbox{\tiny adm.~cuts $C$ of
$t$}}S[P^C(t)]R^C(t)=-t-m[(S\otimes id)\Delta^\prime(t)].
\end{eqnarray}
An alternative formula for the antipode, which one may
easily derive by induction on the number of vertices
\cite{hopf,CK} is
\begin{eqnarray*}
S(t) & = & -\sum_{\mbox{\tiny all  cuts $C$ of
$t$}}(-1)^{n_C}P^C(t)R^C(t),
\end{eqnarray*}
where $n_C$ is the number of single cuts in $C$. This time, we
have a non-recursive expression, summing over all cuts $C$ and
relaxing the restriction to admissible cuts. The overall minus
sign can be incorporated in the sum if we attach an incoming edge
to the root. All cuts which remove this edge are then full cuts,
all the other ones are normal cuts. A Feynman graph corresponding
to a tree with $n$ vertices allows then for $2^n$ cuts. The
$2^{n-1}$ full cuts deliver the counterterm, the $2^{n-1}$ normal
cuts eliminate the subdivergence to deliver the result of the
$R$-bar operation \cite{hopf,CK}.

So far we have established a Hopf algebra on rooted trees, using
the set of rooted trees, the commutative multiplication $m$ for
elements of this set, the unit $1$ and counit $\bar{e}$, the
coproduct $\Delta$ and antipode $S$. We call this Hopf algebra
${\cal H}_R$. Continuing in the manner of \cite{hopf,CK,overl}, we
may label the vertices of rooted trees by Feynman graphs without
subdivergences, in the sense described in the paper and in detail
in \cite{CK,overl} .

Let us also mention that
\begin{equation}
m[(S\otimes id)\Delta(t)]=\bar{e}(t)=0=\sum S(t_{(1)})t_{(2)},
\end{equation}
where we introduced Sweedler's notation $\Delta(t)=:\sum
t_{(1)}\otimes t_{(2)}$, and $id$ is the identity map ${\cal
H}_R\to{\cal H}_R$.
We conclude by defining $\#(t)$ to be the number of vertices
of a rooted tree $t$. This extends to a
monomial of rooted trees in the obvious manner: $\#(\prod_i
T_i)=\sum_i \#(T_i)$.

\subsection{The Hopf Algebra Structure of Graphs and Forest Formula}
The results of \cite{overl} show that for each Feynman graph
$\Gamma$, we obtain a sum of associated rooted tree $T_\Gamma$ and
a coproduct given by
\begin{equation}
\Delta(T_\Gamma)=1\otimes T_\Gamma +T_\Gamma\otimes
1+\sum_{\gamma\subset_X \Gamma}T_\gamma\otimes
T_{\Gamma/\gamma}.\label{copfg}
\end{equation}
Here, $T_\Gamma$ is a sum of rooted trees with decorations,
primitive elements in the Hopf algebra of rooted trees, which are
obtained from Feynman graphs without subdivergences. As the map
$T_\Gamma\leftrightarrow\Gamma$ is one-to-one, we can directly
formulate the Hopf algebra on graphs $\Gamma$, as in section II.

To the coproduct (\ref{copfg}) belongs an antipode given by
\begin{equation}
S(T_\Gamma)=-T_\Gamma-\sum_{\gamma\subset\Gamma}S[T_\gamma]
T_{\Gamma/\gamma}, \label{anti}
\end{equation}
as one straightforwardly checks. Because it is an antipode in a Hopf
algebra of rooted trees, it can be written as a sum over all cuts. Set
$T_\Gamma=\sum_i T_i$ for some decorated rooted trees $T_i$. Then,
\begin{equation}
S(T_\Gamma)
= \sum_i \sum_{\mbox{\tiny all cuts $C_i$ of $T_i$}} (-1)^{n_{C_i}}
P^{C_i}(T_i)R^{C_i}(T_i).\label{anti2}
\end{equation}
Each such cut corresponds to a renormalization forest, obtained
by boxing the corresponding subgraphs in $\Gamma$, and vice
versa \cite{CK}.

Now, let $\phi$ be a ${\bf Q}$-linear  map which assigns to
$T_\Gamma$ the corresponding Feynman integral. Further, let
$\phi_R=\tau_R\circ\phi$ be a map which assigns to $T_\Gamma$ the
corresponding Feynman integral, evaluated under some
renormalization condition $R$. Hence, from $T_\Gamma$ we obtain
via $\phi$ a Feynman integral $\phi(T_\Gamma)$ in need of
renormalization. $\tau_R$ modifies  this Feynman integral, in such a
way that the result contains the divergent part of this
integral. Essentially, $\tau_R$ extracts the divergences of
$\phi(T_\Gamma)$ in a meaningful way \cite{Collins}. Hence, as
$\tau_R$ isolates divergences faithfully, differences
$(id-\tau_R)(\phi(T_\Gamma))$ eliminate infinities in Feynman
integrals. Depending on the chosen renormalization scheme $R$, one
can adjust finite parts to fulfill renormalization conditions. A
detailed study of this freedom from the Hopf algebra viewpoint can
be found in \cite{dkchen}.

We remind the reader of Sweedler's notation:
$\Delta(T_\Gamma)=\sum {T_\Gamma}_{(1)}\otimes {T_\Gamma}_{(2)}$.
Let us consider the antipode $\bar{e}(T_\Gamma)$ in the same notation:
\[
0=\bar{e}(T_\Gamma)=\sum S({T_\Gamma}_{(1)}){T_\Gamma}_{(2)}.
\]
The above map vanishes identically, and it can also be written as
\[
m[(S\otimes id)\Delta(T_\Gamma)]\equiv\bar{e}(T_\Gamma)=0.
\]
But this map gives rise to a much more interesting map, by
composition with $\phi$,
\[
T_\Gamma\to \Gamma_R:=m[(S_R\otimes
id)(\phi\otimes\phi)\Delta(T_\Gamma)],
\]
for it associates the renormalized Feynman integral $\Gamma_R$
\cite{hopf,CK} to the Feynman graph $\Gamma$ represented by a
unique sum of rooted trees.

Its usual definition,
\begin{equation}
\Gamma_R=(id-\tau_R)\left[\Gamma+\sum_{\gamma\subset\Gamma}Z_\gamma
\Gamma/\gamma\right],\label{for}
\end{equation}
is recovered if we define
\begin{equation}
S_R[\phi(T_\gamma)] \equiv Z_\gamma=
-\tau_R(\gamma)-\tau_R\left[\sum_{\gamma^\prime\subset\gamma}
Z_{\gamma^\prime}\gamma/\gamma^\prime\right].\label{usual}
\end{equation}
This map is derived from  the antipode
\begin{equation}
S[T_\gamma]=-T_\gamma-\sum_{\gamma^\prime\subset\gamma}S[T_{\gamma^\prime}]
T_{\gamma/\gamma^\prime}.
\end{equation}
Using $\phi$ to lift this to Feynman graphs, and using the freedom
to alter corresponding analytic expressions according to
renormalization schemes $R$ one ends up with (\ref{usual}).

Note that if one defines
\[
\phi_R=S_R\circ\phi\circ S,
\]
one has $S_R\circ \phi=\phi_R\circ S$ and hence
\begin{equation}
S_R[\phi(T_\gamma)]=\tau_R\left[
-\phi(T_\gamma)-\sum_{\gamma^\prime\subset\gamma}\phi_R(S[T_{\gamma^\prime}])
\phi(T_{\gamma/\gamma^\prime}). \right]
\end{equation}
Hence, in accordance with \cite{hopf,CK,overl} we determine the
$Z$-factor of a graph $\gamma$ as derived from the antipode in the
Hopf algebra of rooted trees. In (\ref{for}), we recovered the
original forest formula in its recursive form. The non-recursive
form is recovered with the same ease, using (\ref{anti2}) instead
of (\ref{anti}) \cite{hopf,CK,overl}.

\newpage

\begin{figure}
\caption{A Feynman graph, its divergent subgraphs, its forests,
and the corresponding tree with appropriate full and normal cuts.
>From this one calculates the antipode and the renormalized Green
function.}
\end{figure}

\begin{figure}
\caption{Forests and counterterms as cuts and antipodes. Each
forests (thick grey lines) contains an analytic expression which
has to be evaluated using $R$.}
\end{figure}

\begin{figure}
\caption{Vertex weights and tree factorials. For the tree $t$
given at the left in the upper row we indicate the vertex weights
attached to each vertex. We also give the Feynman diagram
corresponding to a chosen common one-loop self-energy decoration
at each vertex.}
\end{figure}

\begin{figure}
\caption{Generalized vertex weights and tree factorials
which can appear if there are decorations of arbitrary loop order.
The factorial of this decorated tree is $2\times 5=10$.}
\end{figure}

\begin{figure}
\caption{A Yukawa self-energy at three loops.}
\end{figure}

\begin{figure}
\caption{Examples of Feynman graphs, their decorated rooted trees
and the corresponding matrix calculus. From top to bottom, the
bare function $\Gamma(t)$, evaluates for the indicated decorated
rooted trees to ${\bf M}_1{\bf v}_0 z^{-x}$, ${\bf A}_1{\bf
M}_2{\bf v}_0 z^{-2x}$, ${\bf A}_1{\bf M}_2{\bf M}_3{\bf v}_0
z^{-3x}$.}
\end{figure}

\begin{figure}
\caption{The action of $B_-$ on a rooted tree.}
\end{figure}

\begin{figure}
\caption{The action of $B_+$ on a monomial of  trees.}
\end{figure}

\begin{figure}
\caption{An elementary cut $c$ splits a rooted tree
$t$ into two components, the fall-down $P^c(t)$ and the piece which
is still connected to the root, $R^c(t)$.}
\end{figure}

\begin{figure}
\caption{An admissible cut $C$ acting on a tree $t$, leading to a
monomial of trees. One of the factors, $R^C(t)$, contains
the root of $t$.}
\end{figure}

\begin{figure}
\caption{The coproduct, worked out for the trees
$t_1,t_2,t_{3_1},t_{3_2}$, from top to bottom. In the last line we
give one full and all normal cuts of the tree.}
\end{figure}

\newpage
\setcounter{figure}{0}

\bookfig{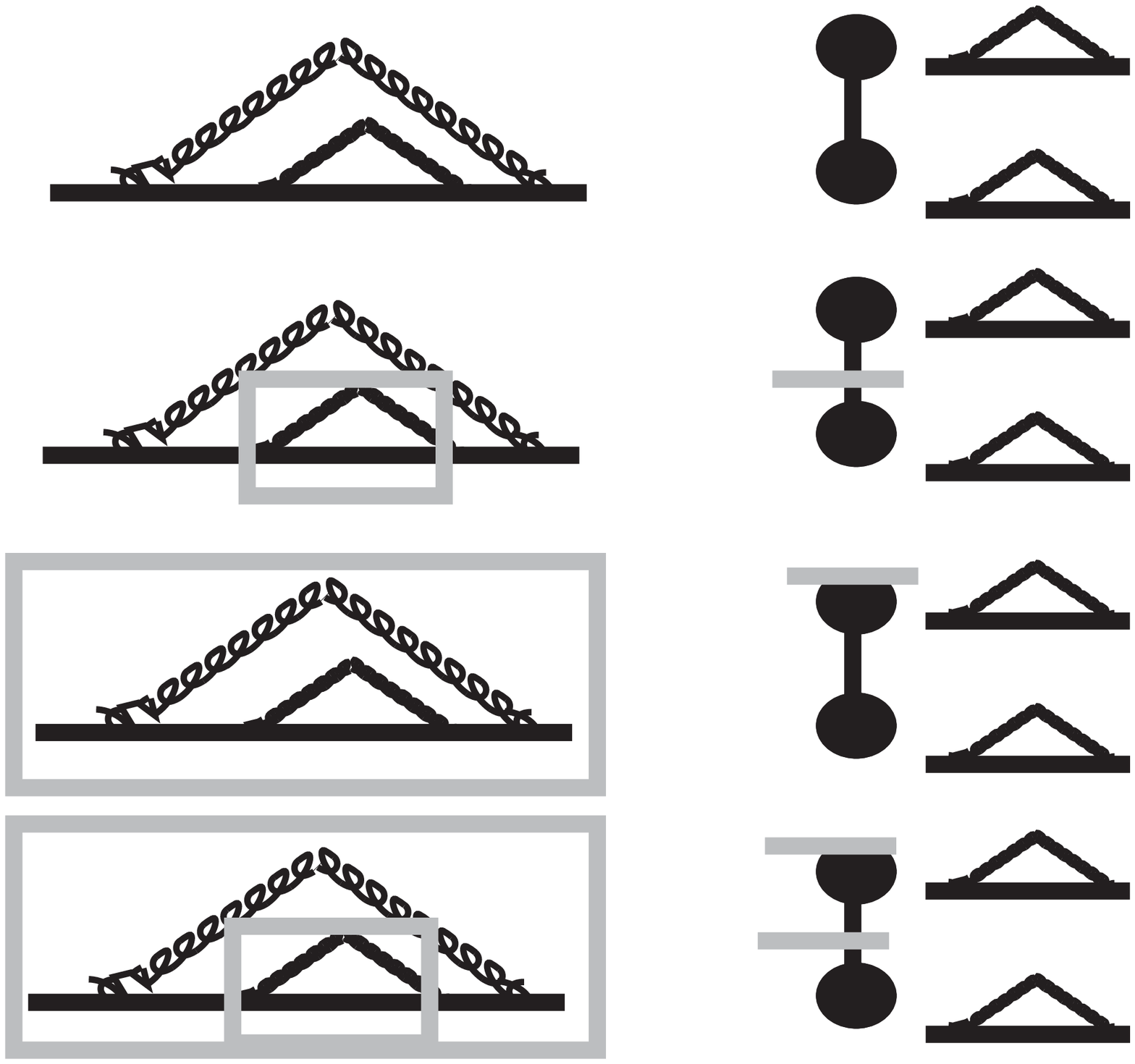}{ex}{b1}{}{6}

\bookfig{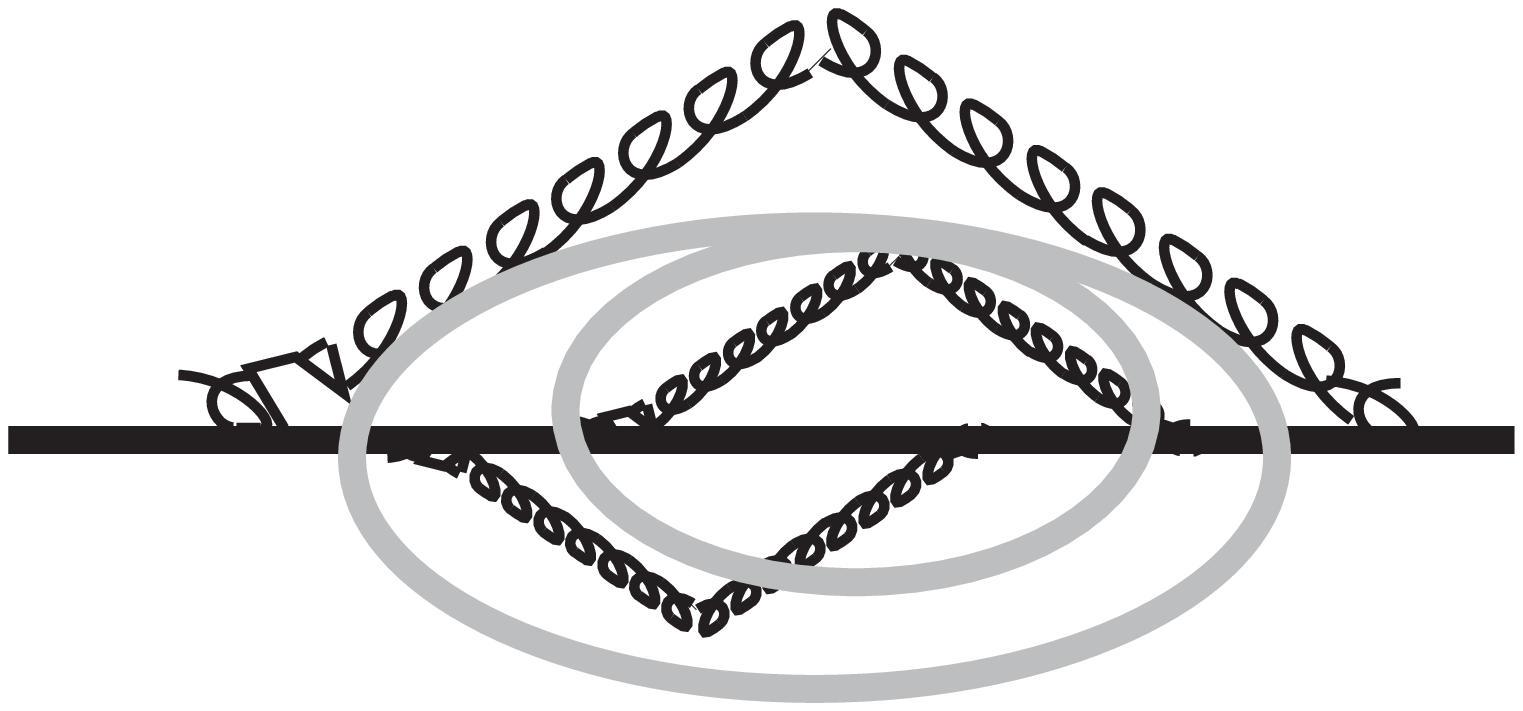}{Ex}{b2}{}{4}

\bookfig{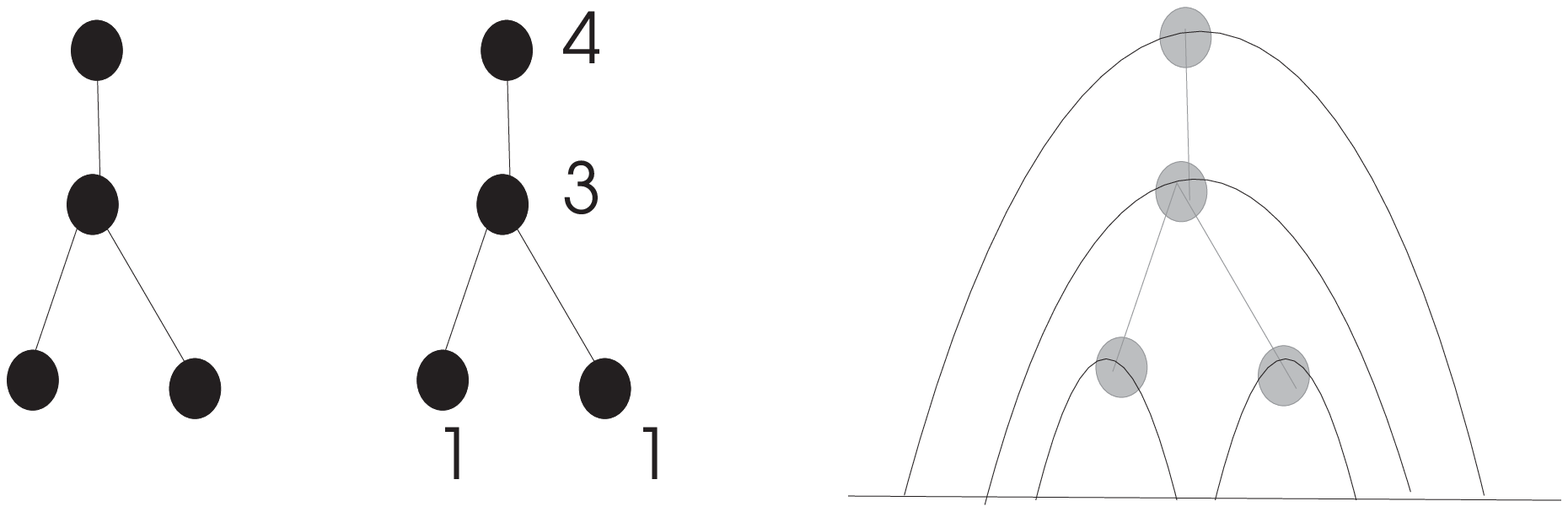}{tf}{f3}{}{4}

\bookfig{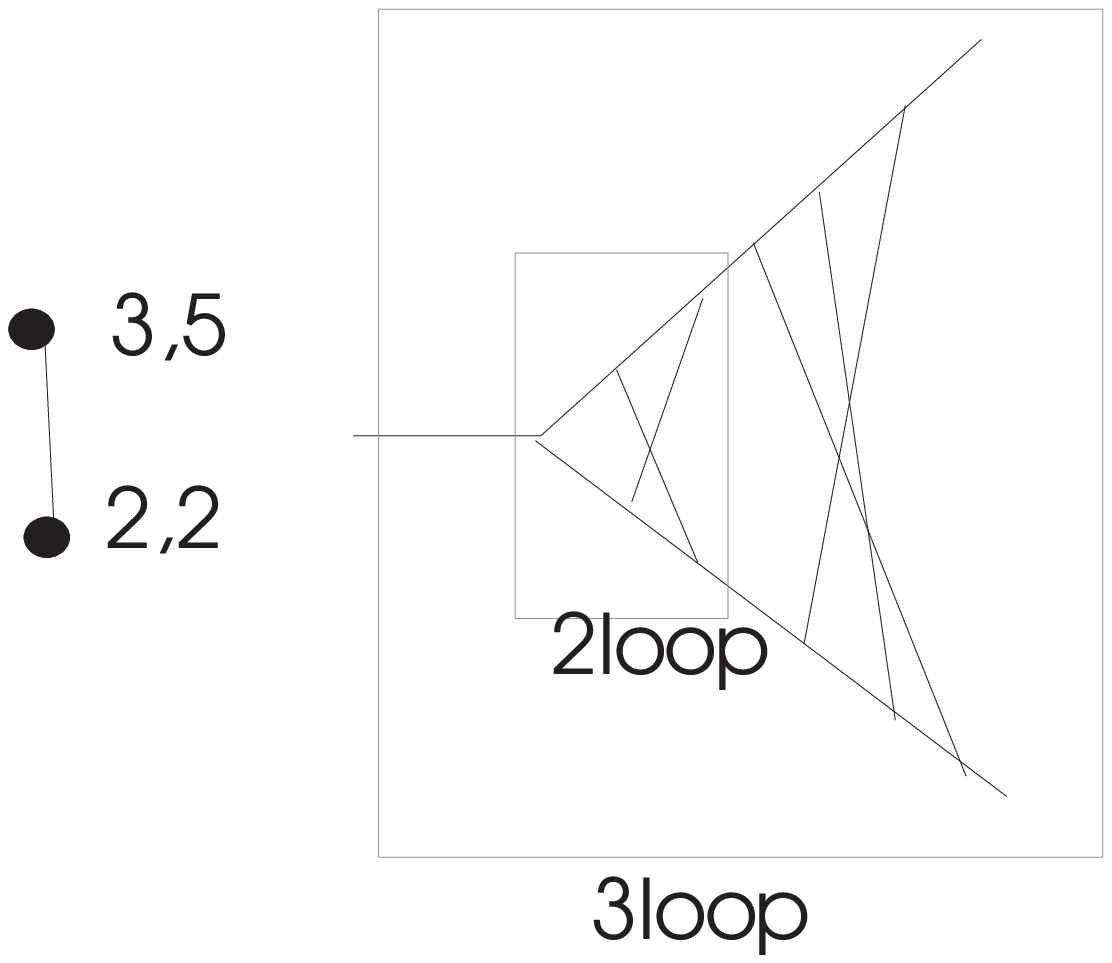}{!}{gvw}{}{5}

\bookfig{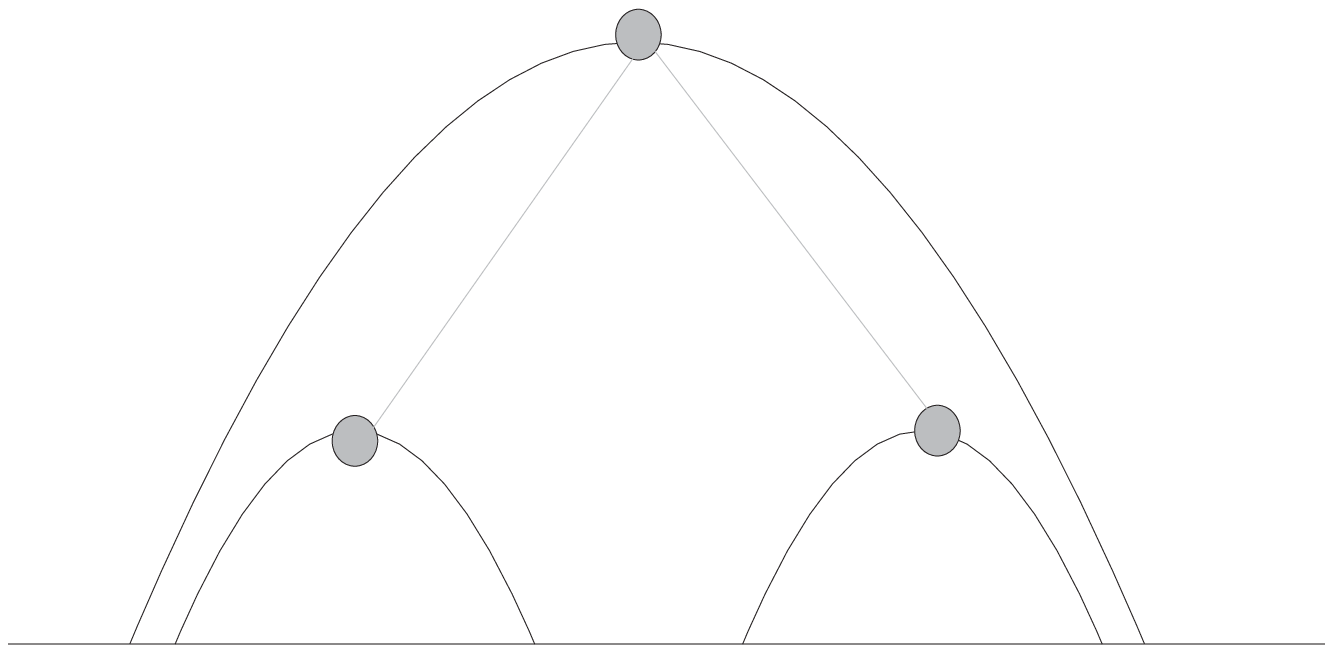}{!}{rain}{}{4}

\bookfig{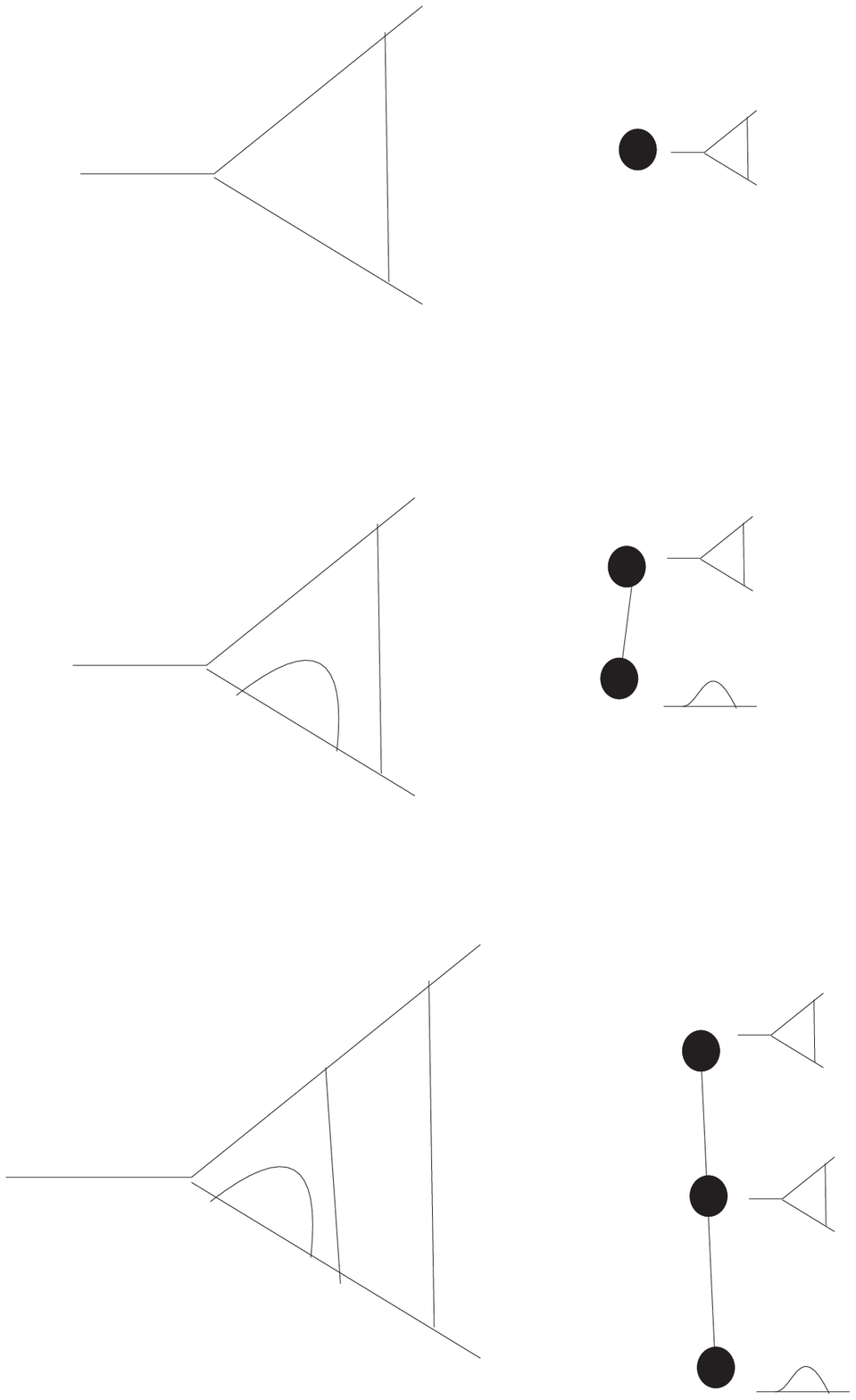}{!}{matrix}{}{7}

\bookfig{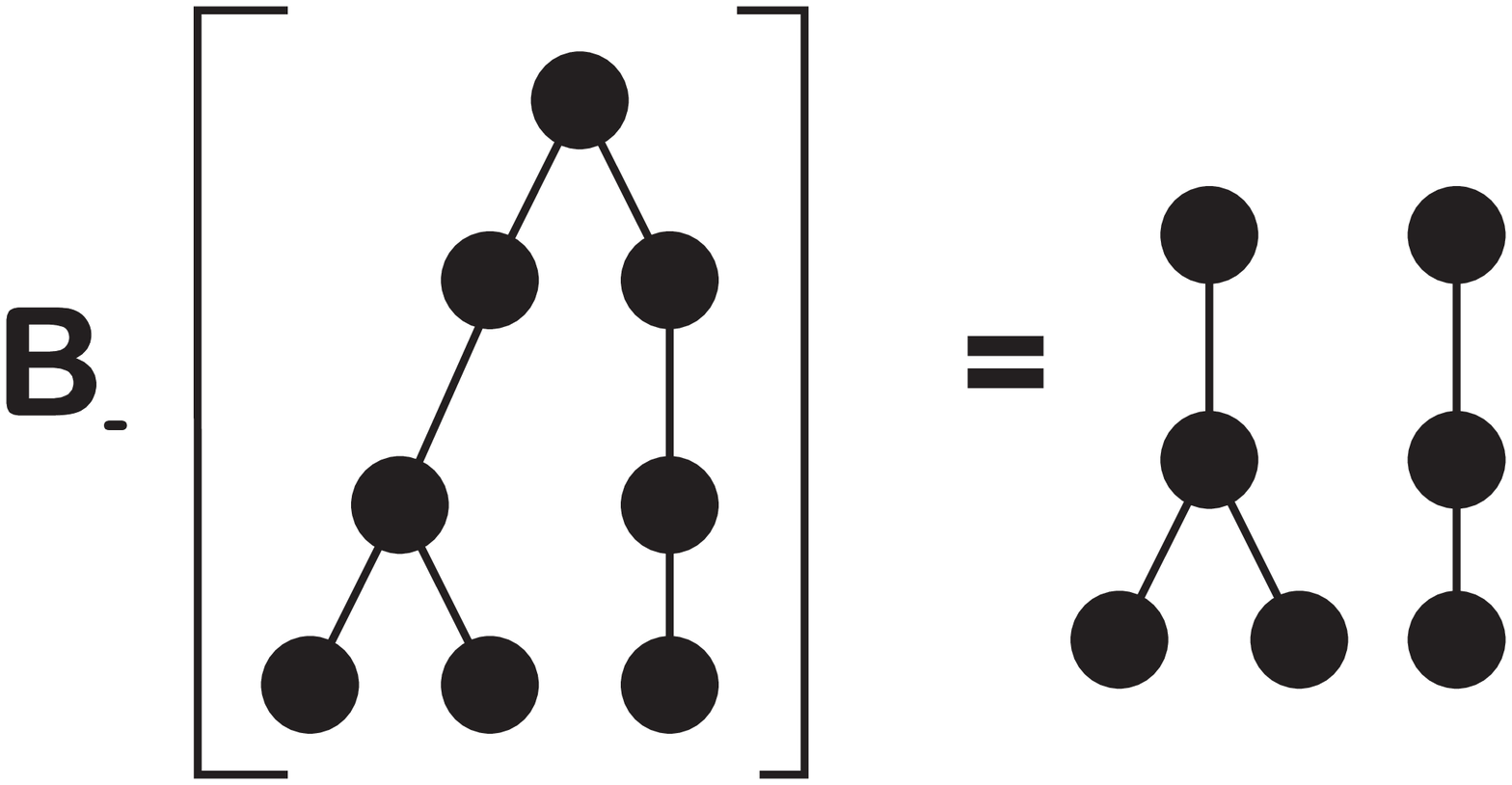}{$B_-$}{guill-}{}{3}

\bookfig{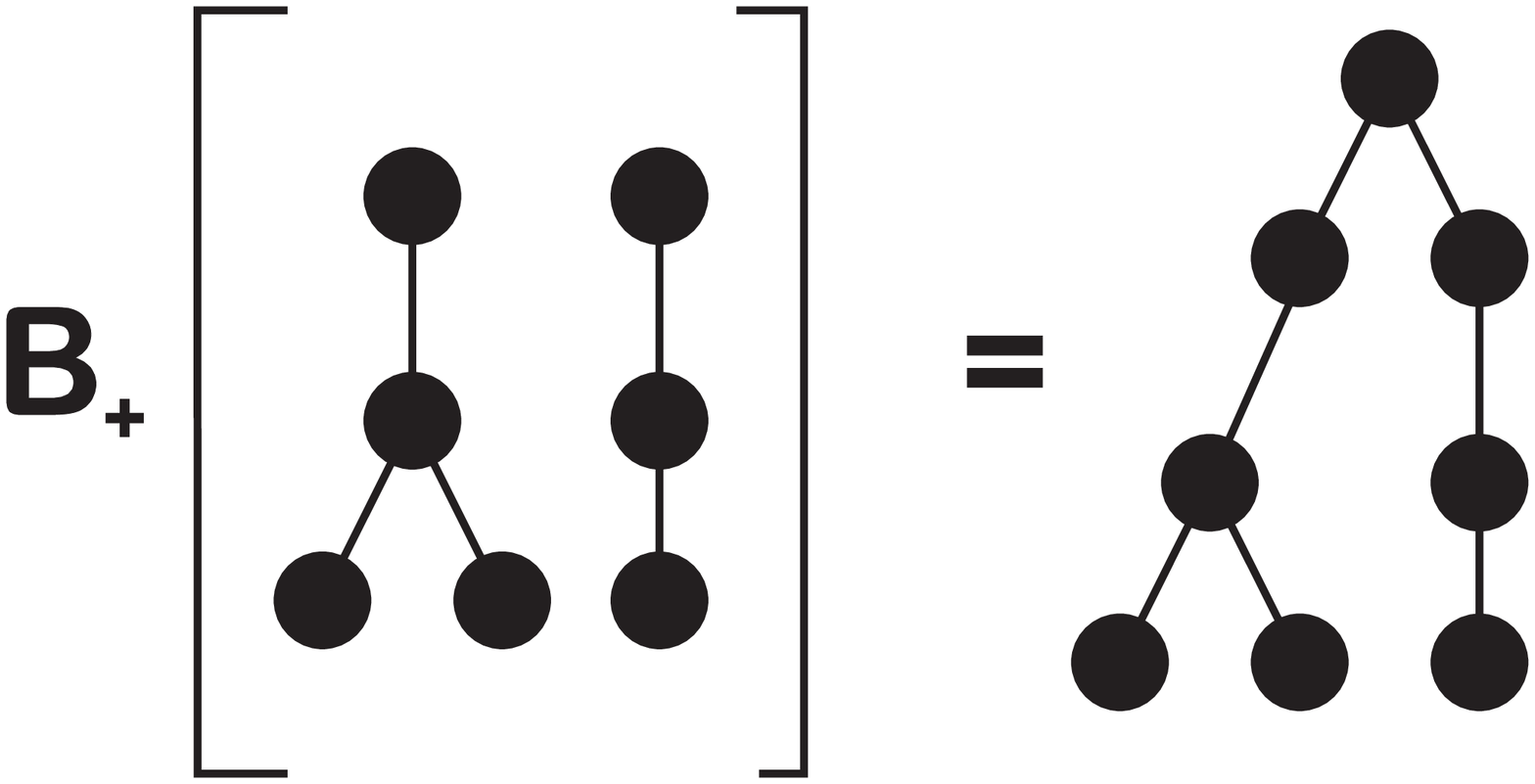}{$B_+$}{guill+}{}{3}

\bookfig{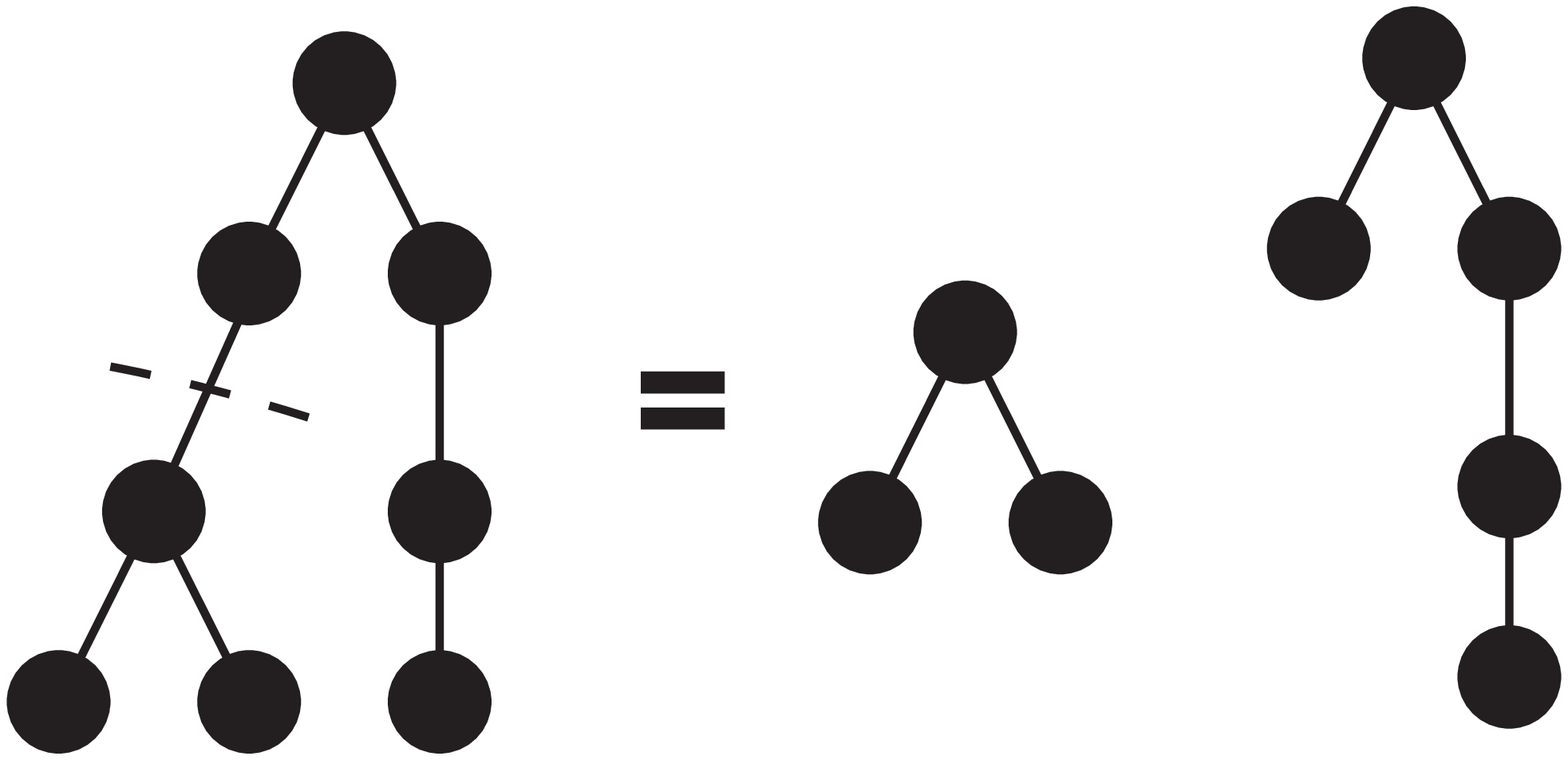}{Elementary cut.}{ecut}{}{3}

\bookfig{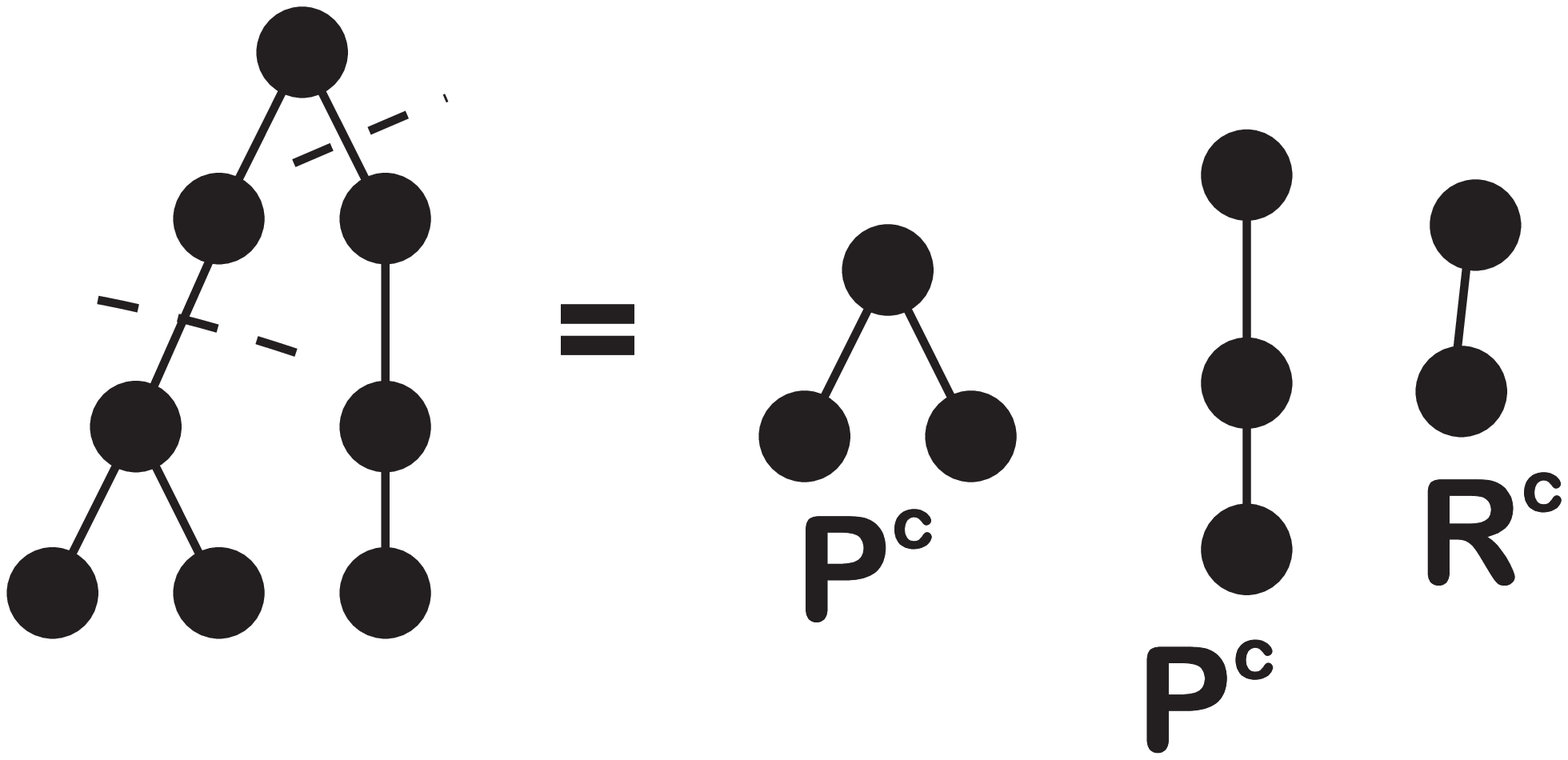}{An admissible cut.}{cut}{}{4}

\bookfig{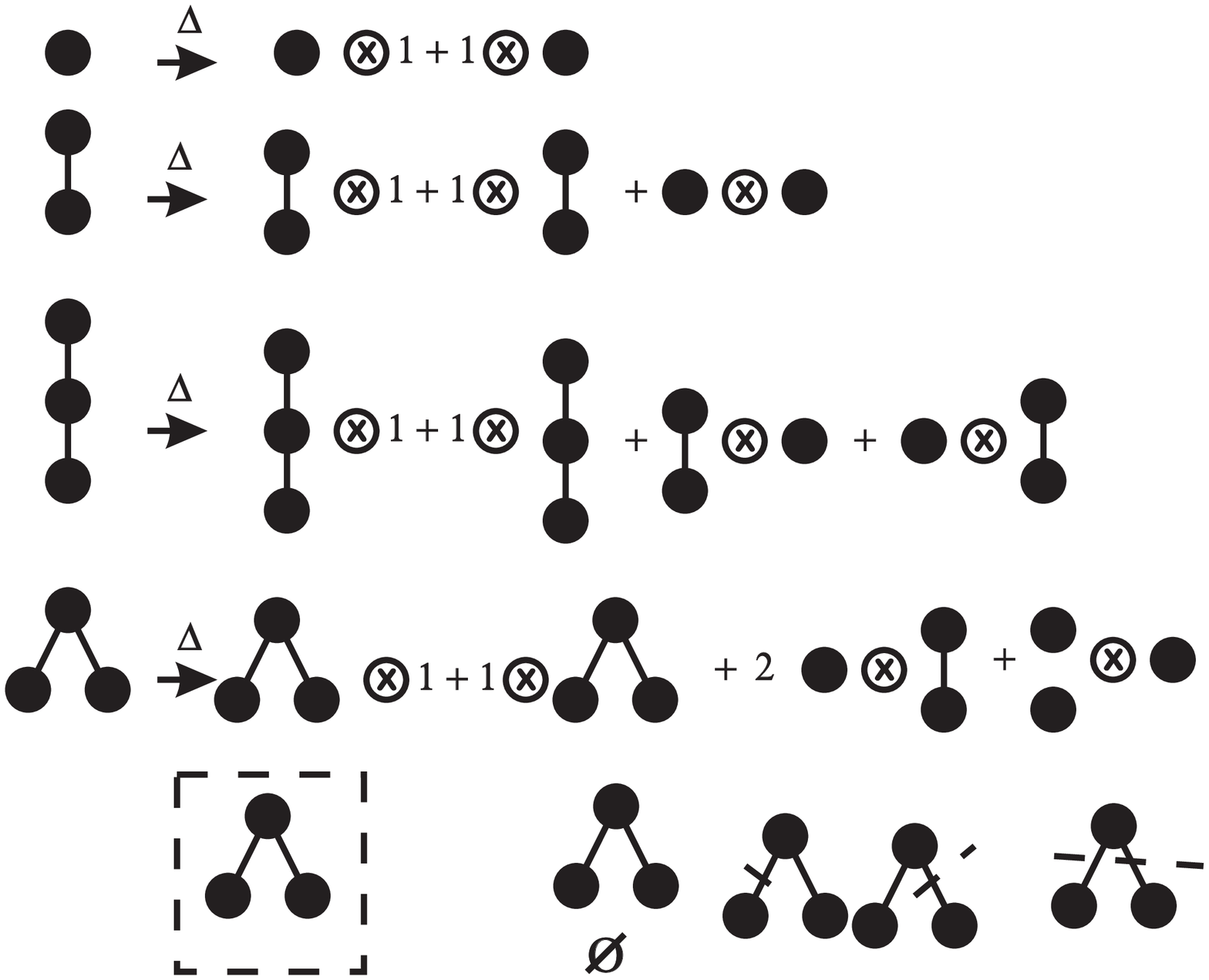}{The coproduct.}{cop}{}{6}

\end{document}